# Music and art – a study in cross-modal interpretation

Paul Warren, Paul Mulholland, Naomi Barker
{paul.warren, paul.mulholland, naomi.barker}@open.ac.uk
The Open University, U.K.

## ABSTRACT

Our study has investigated the effect of music on the experience of viewing art, investigating the factors which create a sense of connectivity between the two forms. We worked with 138 participants, and included multiple choice and open-ended questions. For the latter, we performed both a qualitative analysis and also sentiment analysis using text-mining. We investigated the relationship between the user experience and the emotions in the artwork and music. We found that, besides emotion, theme, story, and to a lesser extent music tempo were factors which helped form connections between artwork and music. Overall, participants rated the music as being helpful in developing an appreciation of the art. We propose guidelines for using music to enhance the experience of viewing art, and we propose directions for future research.

## 1 Introduction

The study of the relationship between visual art and music has a long history; we give a brief outline in subsection 1.1. As we shall see, this has encompassed both the relationship between elemental visual and auditory experiences, and also the relationship between more complex experiences. Our interest in this paper is with the latter. We started from the view that an important function of both art and music is to convey and invoke emotions; thus we expect that shared emotions will form an important connection between artworks and music. However, we know from the literature and from our own previous work (Mulholland et al., 2023) that other links may help form relationships between the visual and auditory, e.g. a theme or an idea, a story, or a shared sense of movement; again, we discuss this in more detail later. Our goal is to understand better the links which people make between art and music. Our objective is to use music to enhance the experience of viewing art. However, we note that there are other potential applications, e.g. Park et al. (2024) were interested in using images to retrieve music through shared moods and themes.

In our study participants viewed an artwork, first without musical accompaniment, and then with four one-minute musical extracts. On each occasion, we asked the participants about their experience of the artwork, and how that was influenced by the music. Our study included a quantitative analysis of participants' responses, e.g. participants were asked to rate the degree of connectivity between artwork and each piece of music; and also a qualitative analysis of their spoken or written opinions. We have supplemented this qualitative analysis by using text processing techniques, specifically sentiment analysis. A subsidiary goal of our work was to determine to what extent such techniques confirmed a more conventional qualitative analysis, and to what extent these techniques added further insight.

In the remainder of this introductory section, subsection 1.1 provides an overview of related work; subsection 1.2 describes a previous pilot study; subsection 1.3 explains our framework for describing emotions, in particular the valence-arousal plane; and subsection 1.4 describes our research questions. Section 2 then explains our methods, including how we chose the artworks and musical extracts, and a description of the questions we asked participants. Section 3 presents our results. Section 4 discusses some implications of our study. Finally, Section 5 draws some conclusions.

## 1.1 Related work

We begin, in subsection 1.1.1, by saying something about aesthetic appreciation generally; and then, in subsection 1.1.2, we review the literature on the relationship between art and music.

### 1.1.1 Aesthetic appreciation

The last few decades have seen an evolving scientific literature on aesthetic appreciation and aesthetic development. An early researcher was Housen (Housen, 1983, 1987; DeSantis & Housen, 2009), who used stream-of-consciousness interviews, to develop a theory of aesthetic development involving five stages. These range from stage I, where the viewer is making concrete observations, to stage V, where the viewer is aware of the background to the work and " combines personal contemplation with views that broadly encompass universal concerns". The emphasis here is cognitive. Indeed, Housen was a co-founder of *Visual Thinking Strategies*[1], an organisation which is concerned with using visual art to encourage engagement in, and enjoyment of, learning generally.

Whilst Housen's research was interview-based, and her goal was to encourage the appreciation of art and develop critical thinking skills generally, other research has been laboratory-based with more purely scientific intentions. Leder et al. (2004) developed an information-processing model of the aesthetic experience, which Leder and Nadal (2014) reviewed ten years after its initial publication. The model proposed an interaction between cognitive and affective processing; the former leading to aesthetic judgement, the latter to aesthetic appreciation. In their review, Leder and Nadal (2014) further stress the importance of this interaction. They see aesthetic experience as drawing on psychological processes relating to a range of other human activities, "ranging from economic decision making to empathy". At the same time as calling for increased neurobiological understanding, they also call for progress in "our understanding of the cognitive and affective processes that enable us to create and appreciate art and aesthetics". Although differing in their intentions and in their methods and their goal, Housen (Housen, 1983, 1987; DeSantis & Housen, 2009) and Leder et al. (2004) share the view that cognitive processing is an important component of aesthetic experience.

Music, if it is the right music, can be regarded as setting a context for an artwork; the most obvious other forms of context are the title and any additional information. Leder et al. (2006) looked at the effect of titles, comparing 'descriptive' against 'elaborative' titles. For example, for a picture by Cezanne originally titled *Mont Saint Victoire*, the descriptive title was *Mountain*, whilst the elaborative title was *Different proportions*. They found that, for abstract art but not representational art, titles could improve understanding. Moreover, for abstract art, elaborative titles could improve understanding more than descriptive titles. However, the effect of elaborative titles required a certain amount of time, e.g. ten seconds. When participants were provided with the title for only one second, the reverse was true, i.e. the elaborative titles led to less understanding than descriptive titles. They also found, in an experiment using abstract paintings, that judgements concerning liking were made faster than judgements concerning understanding. Leder et al. (2006) were in part concerned with validating their model of aesthetic appreciation (Leder et al., 2004; Leder & Nadal, 2014). From the standpoint of our study, what is interesting is the distinction they make between liking and understanding. From the standpoint of encouraging aesthetic appreciation, they could be criticized on the grounds that the elaborative titles are likely to represent the subjective views of the experimenters, not those of the artists, and in any case inhibit the viewers from forming their own views.

---

[1] https://vtshome.org/ A similar goal is shared by the *Artful Thinking* programme (http://pzartfulthinking.org/), developed by *Project Zero* (https://pz.harvard.edu/) at the Harvard Graduate School of Education.

Russell (2003) considered the effect of providing additional information, beyond title and artist. He makes a distinction between meaningfulness and hedonic value, or pleasingness. The importance of meaningfulness in the aesthetic experience has been recognised for at least 50 years, and Russell (2003) provides a brief introduction to the topic. He reports a between-participants experiment with three conditions: provision of title and author; provision of title, author, plus additional information; and a control condition with the provision of no information whatsoever. For four out of twelve paintings the provision of the additional information led to a significant increase in meaningfulness over the control condition. For only one of the paintings did the additional information make a significant difference to pleasingness, and in this case the pleasingness was significantly reduced over that for title and author alone[2]. The authors hypothesized that the inability for most paintings to detect a difference in pleasingness was due to the between-participants experimental design. In a second experiment, they used a within-participants design, which is in principle more sensitive. In this experiment, the additional information did result in a significant increase in pleasingness. Russell (2003) interprets this increase in pleasingness as resulting from the sense of "making a successful interpretation" of the painting; the small increase arising from the fact that interpretation is just one of a number of factors affecting pleasingness. From our standpoint, an additional finding from the first experiment is particularly important; there was no significant correlation between meaningfulness and pleasingness. This suggests that meaningfulness and pleasingness need to be treated as independent factors.

Ferraris et al. (2023) looked at the provision of context in the presentation of objects digitally. Specifically, they worked with objects from Ancient Egypt, i.e. objects primarily of historical interest rather than purely aesthetic interest. Provision of context included providing background scenery and a human figure. They found that contextual information provided a sense of scale and improved the perceived realism of the digital object. More relevant to our study, they also found that contextual information supported the generation of a narrative.

### 1.1.2 Music and art cross-modality

Research on the relationship between art and music goes back at least to the first half of the 20th century. Cowles (1935), for example, was concerned with how study participants associated classical music extracts and artworks. He found "partial agreement" between participants in the selection of matching music and artworks. In parallel experiments, Cowles (1935) used two different questions: asking in one case for a combination where "affective mood [was] most similar", and in the other to select the picture which "best fits the music". In the latter case, an important factor in the selection was the degree of dynamism in the artwork and the music. Pictures with "content capable of motor activity were nearly always selected with the musical selections of prominent dynamic changes"; whilst "pictures of slight content were nearly always selected with music of relatively weak dynamic qualities". When asked for combinations with similar affective mood, participants' introspective reports mentioned marked rhythm, changes in loudness, peculiar tempo. When asked for a picture which best fits the music, participants mentioned mood, colour and lines. Cowles (1935) study made use of both musical and unmusical participants; he found no difference between the two groups.

---

[2] This was a somewhat exceptional case. The painting in question was *Burial* by George Grosz. As Russell (2003) notes, the description explained "that the painting was produced after the artist had been committed to a psychiatric hospital with a mental breakdown following having witnessed appalling casualties while fighting in the First World War". This additional information may have led to a reduced sense of pleasingness.

Much later, Limbert and Polzella (1998) used impressionist and abstract paintings and two pieces of classical music which might be considered to match these two categories. Participants were divided into three groups: those who viewed paintings and listened to matching music; those who viewed paintings with unmatched music; and those who viewed paintings with no music. In general, participants found the impressionist paintings more beautiful and the abstract paintings more active. This effect was increased when the paintings were viewed with matching music. It was also the case that both groups of paintings, when viewed with impressionist music, were regarded as more beautiful and less active than when viewed with abstract music.

Palmer et al. (2013) also used classical music, but for the visual stimuli they used more elemental stimuli, specifically colours. They found that "faster music in the major mode produced color choices that were more saturated, lighter, and yellower", whereas "slower, minor music produced … choices that were desaturated, darker, and bluer". Participants were then asked to rate the musical extracts and colours along four emotion scales (happy/sad, angry/calm, strong/weak, lively/dreary). There was found to be a strong correlation between the scores on the four emotion scales for the musical extracts and the scores for the colours which participants had associated with each musical extract. To support their view that these associations were mediated by common emotional associations, they undertook two further experiments using a selection of faces representing emotions. In one experiment, participants were asked to associate the colours with faces; in the other experiment participants associated the musical extracts with faces. The results of these experiments were consistent with the previously-found relationship between music and emotion, and artwork and emotion.

Further support for the emotional mediation theory came from Whiteford et al. (2018). Like Palmer et al. (2013) they used musical extracts for the audio stimuli, although from a wider range of genres, and colours for the visual stimuli. Whiteford et al. (2018) found that the correlation between music and colours became insignificant after removing the emotional effects. Moreover, a reduction of ten emotional factors to two, using parallel factor analysis resulted in dimensions which were consistent with the dimensions of arousal and valence; the latter term is equivalent to pleasantness, e.g. as used by Russell (1980). We discuss these two dimensions further in the next subsection. It is also worth noting that Whiteford et al. (2018) were careful to exclude synaesthetes from their study. The phenomenon we are discussing here, and which we investigate in our study, is quite distinct from synaesthesia. Unlike synaesthesia, cross-modality is experienced by the majority, if not all, of the population.

Albertazzi et al. (2015), using both complex visual and audio stimuli, came to conclusions which conflict with a theory of association purely based on emotion. They investigated the cross-modal associations between artworks by the Italian artist Matteo Boata (https://www.matteoboato.net/) and classical Spanish music. They found that the musical modes normally associated with happiness and sadness, i.e. major and minor, did not play a significant role in associating the two art forms. Rather, the main perceptual features responsible for the associations were "hue, lightness, timbre, and musical tempo". They also found "no substantial difference … between expert and non-expert" participants.

Iosifyan et al. (2022) investigated the correspondence between paintings, both figurative and abstract, and sounds. This time the sounds were not musical, but consisted of embodied sounds, e.g. the sound of steps on a wooden floor and the sound of human drinking; and synthetic sounds, e.g. electronic music samples and NASA space sounds. They found that embodied sounds were more strongly associated with figurative paintings, and synthetic sounds with abstract. Their interest was in how sound affects aesthetic appreciation, perceived meaningfulness and immersive experience of paintings. These effects were gauged by asking a direct question (e.g. "Is this painting meaningful") to which participants responded on a six point scale. Their conclusion was that congruence between sound and painting, as determined by an initial pre-test, increased all three of these qualities; whilst sound embodiment increased immersive experience of paintings. The inclusion of the meaningfulness question is particularly interesting. It is not apparent that any explanation was given of the term; presumably participants were free to interpret meaningfulness in whatever way they wished.

In all the studies described so far, participants were required to be relatively passive, e.g. assessing how beautiful or active a painting seemed, or associating a piece of music with a colour. However, other studies provided participants with a more active role. Fosh et al. (2013) were interested in the relationship between music and sculpture. Participants in their study were able to walk freely through a sculpture garden, although guided by 'trajectories', listening to different musical extracts at each sculpture. They were subsequently interviewed to gauge their reaction to the experience. These interviews revealed how the music affected the appreciation of the sculptures, e.g. one participant was apprehensive about approaching a statue because of the associated eerie music. A similar, later experiment based in a sculpture park confirmed the influence of music on the appreciation of art; amongst a number of comments to this effect, one participant said "The overwhelming thing for me was how it made it a different experience …" (Hazzard et al., 2015).

Spence (2020b) reviews a range of studies investigating cross-modal correspondences; in particular he includes studies involving complex stimuli. He makes the point that, in these studies, the musical extracts were professionally composed with the intention of eliciting an emotional response. Hence it is not surprising that the emotional mediation theory has so much explanatory validity for complex auditory stimuli. He makes the further point that, just because there is a correspondence between music and visual stimuli, that does not mean that one stimuli will influence the other when presented together. Spence (2020b) completes his review by examining four potential explanations for cross-modal correspondences:

- Statistical, i.e. reflecting statistical regularities in the environment.
- Structural, i.e. "based on common principles of neural encoding (or representation)"; suggesting this may account for intensity-based correspondences.
- Semantic, or linguistic, i.e. resulting from the common terms used for different stimuli. Whilst his example is 'high' and 'low' used for pitch and spatial elevation, there may also be examples relating to art and music, e.g. 'busy'.
- The emotional mediation account, which we have already discussed, and which he suggests may have more importance when discussing complex stimuli, e.g. music and art. Counter to this, we have already seen (Palmer et al., 2013; Whiteford et al., 2018) that emotional mediation has been proposed in situations where the visual stimuli were elemental.

A longer paper by the same author (2020a) reviews the support for the emotional mediation theory to explain correspondences between audio and visual stimuli, concluding that there is support for emotional mediation as "one of the key factors". That said, the author asks whether we need the emotional mediation account, but can rather use an account (the semantic differential account) based not on specific emotions but on the three dimensions of valence, arousal and dominance, which some authors use to describe emotions, and which we describe in the next subsection. In practice, there seems little difference between the two approaches since, as Spence himself notes, individual emotions can be mapped into this space. Finally, one additional and important point which Spence (2020a) notes is the difference between induced (or felt) and perceived emotions; the former being felt by the listener, while the latter being an emotion associated with the stimuli but not necessarily felt by the listener. Zentner et al. (2008) discuss this distinction in some detail.

Özger and Choudhury (2023) were interested in the effects of the provision of information and music whilst viewing art. For artworks, they used Goya's Black Paintings; for music they used the first movement of Dimitri Shostakovich's First Symphony, chosen for its "macabre tone", i.e. for its emotional alignment with the paintings. They divided their participants into three groups: those who were provided with information about the paintings; those who listened to music whilst viewing the paintings; and those who were both provided with information and listened to music. They found that the group provided with both information and music was significantly more emotionally affected than the group provided with information alone. The emotional affect of the group provided with music alone was greater than that for information alone, and less than that for music and information; however these differences were not significant. Moreover, mirroring the comment of Cowles (1935) about musical training, they found that knowledge and interest in art had no effect on the emotional impact.

Janzen et al. (2023) were concerned with how background music would affect the experience of visual art, specifically when viewed naturally in a museum. Participants viewed an abstract painting by Kandinsky, chosen because of its "lack of semantic elements". Participants were divided into five groups: those who viewed without music; and those who viewed with happy, scary, peaceful, and sad music played on headphones. In choosing these emotional categories, they made use of categorization of emotions according to the dimensions of valence and arousal. Specifically, the four categories of music represent the four quadrants of the valence-arousal plane. Valence represents the pleasantness of an experience, whilst arousal represents the experience's intensity; we explain these terms in more detail in subsection 1.3. For the music, they chose four one-minute passages from the operas of Wagner. For each participant, the chosen passage was repeated throughout the period the participant viewed the artwork. They found that positively-valenced music led to the artwork being assigned a significantly more positive valence than in the absence of music.

Isaacson et al. (2023) were also interested in how music can enhance the appreciation of art, or create an enhanced multi-modal experience. They chose three paintings, representing a range of content, and for each painting three pieces of music which they felt matched the painting. Participants were invited to choose a painting and respond to the three questions: 'what do you see', 'what do you think', and 'what do you ask'. They also completed five, five-level semantic differential scales; thereby indicating their views on the painting's excitement, beauty, calmness, happiness and likability. They then listened to the three pieces of music and chose the one which most fitted the painting. For this particular piece, they were then asked to complete the five semantic differential scales. Subsequent analysis revealed that, calculated over all the participants' responses, for each scale there was a significant positive correlation between the rating for the painting and the rating for the music chosen as most fitting. Finally, participants were asked again the three questions about seeing, thinking and asking, plus why they had chosen this artwork. They found that the music not only intensified the experience, but also led some participants "to perceive dynamic movement in the static image and\or attribute self-expressive associations to it". In the context of our study, it is interesting to read how the experimenters chose the three pieces of music for each painting (Isaacson et al., 2023). The explanation given is relatively light on musical theory, or indeed any theoretical framework, and more concerned with the feelings expressed in the music, e.g. one piece of music is said to evoke "an association of contemplation and waiting for what is to come".

We draw two general conclusions from all this preceding work. Firstly, art and music cross-modality is a real effect, and music can be used to influence the appreciation of art. Secondly, whilst emotion may be a dominant factor in some situations, e.g. when one of the modalities is elemental (Palmer et al., 2013; Whiteford et al., 2018), there are other factors which influence a sense of connection between art and music.

## 1.2 Pilot study

In what can be regarded as a pilot for our current study, Mulholland et al. (2023) were interested in increasing the appreciation of art through 'citizen curation', i.e. encouraging visitors to add their own commentaries to works of art. They were further interested in the extent to which this would be enhanced by music. They observed that not only were some participants making an emotional connection between artworks and music, but also they were finding a narrative connection, and a connection through a sense of movement in the artwork, i.e. between the imagined movement in the picture and the rhythmic quality of the music. Although not discussed in their paper, a subsequent analysis revealed a fourth form of connection, through a shared non-narrative theme. Regarding the narrative connection, the importance of this in aesthetic appreciation has been identified by Housen (2002), who regards the construction of a narrative a feature of the first stage of her five stage categorization of aesthetic development.

## 1.3 Describing emotions – the valence-arousal plane

A variety of schemes have been developed to categorize emotions. Much of this work is generic, and we describe some of this later in this subsection. However, we begin by looking at work specifically targeted at emotions associated with music and art. A well-known example of the former is the Geneva Emotional Musical Scale (GEMS), developed by Zentner et al. (2008). Zentner et al. (2008) were specifically concerned with induced emotions, although in principle we see no reason why the scale could not be used for perceived emotions. GEMS has 9 scales: wonder, transcendence, nostalgia, tenderness, tranquility, joy, power, tension, sadness. Each scale has a number of associated labels, i.e. associated adjectives; making 45 labels in all[3].

More recently, Christensen et al. (2023) have created a schema to describe the emotions associated with artworks. In fact, they have developed two sets of terms: descriptive terms, i.e. relating to how an artwork looks; and impact terms, relating to how an artwork makes viewers think and feel. They used scholars to create the terms, and then study participants to generate words for each term. This enabled them to compute semantic similarity between terms and use network analysis to reduce the descriptive terms to five dimensions, and the impact terms to four dimensions. Two of these impact dimensions contain terms related to positive and negative affect, including variations in arousal. The other two dimensions are concerned with immersion, e.g. *enraptured*, *interested*, and transformation, e.g. *edified*, *enlightened*, *inspired*.

Because we wanted a framework relevant to both art and music, for our work we have used a schema from the generic psychological literature on emotion. This literature is extensive, Russell (1980) provides an overview. During the last century, evidence arose that the range of human emotions could be described in terms of a limited number of independent dimensions. Russell and Mehrabian (1977) investigated this hypothesis, using the dimensions of:

- pleasure, i.e. "from extreme pain or unhappiness … to extreme happiness or ecstasy";
- arousal, i.e. "from sleep … to frenzied excitement";
- dominance, i.e. "from feelings of total lack of control or influence … to feeling influential and in control"

In two studies, Russell and Mehrabian (1977) showed that these three dimensions were necessary and sufficient to describe a wide range of emotional terms, i.e. the three dimensions accounted for most of the variance in use of the emotional terms.

---

[3] The current GEMS is described at https://musemap.org/resources/gems, and differs slightly from that described in Zentner et al. (2008). There are, in fact, three versions of GEMS. Besides the one described above, there are two shorter scales.

Russell (1980) focused on the first two of these dimensions, i.e. ignoring dominance. He developed a circumplex model of affect, in which eight affect concepts could be displayed in circular order in a two-dimensional plane, with axes corresponding to the first two dimensions described above, see Figure 1. He was also able to locate 28 words relating to emotion in a similar fashion, circularly around the plane. In our work, we have used this two-dimensional framework to describe artworks, musical extracts, and study participants' descriptions of their experience. We use just the two dimensions of pleasantness and arousal for experimental simplicity. We justify this on the grounds that, whereas Russell and Mehrabian (1977) found only a correlation of 0.03 between pleasure and arousal, they found a correlation of 0.40 between pleasure and dominance and of 0.15 between arousal and dominance. However, there is a loss of expressivity in ignoring the dominance dimension; Russell and Mehrabian (1977) point out that, without taking account of dominance, it is impossible to separate, e.g. angry from anxious or alert from surprised. We return to this in Section 4. Following later authors, we use the term 'valence' to describe the pleasantness-unpleasantness axis. Thus, we imagine emotions, both incited and perceived, as occupying positions in the valence-arousal (VA) plane.

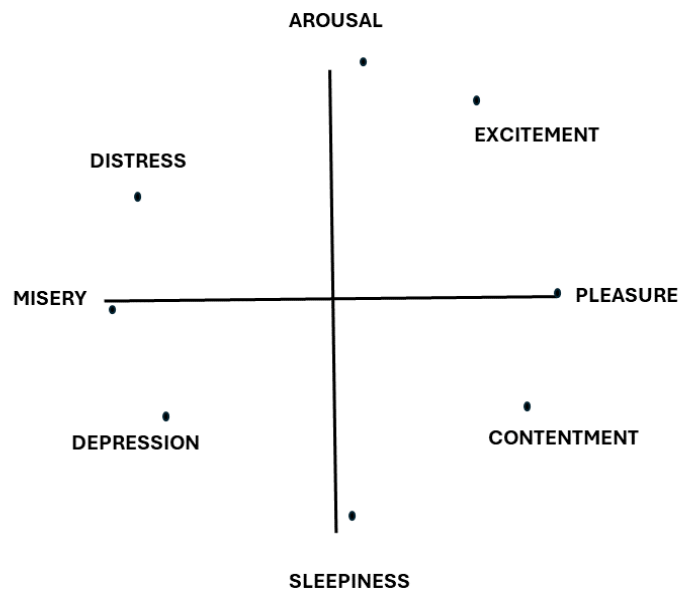

Figure 1. Eight affect concepts, in a circular order, positioned by pleasure (horizontal dimension) and arousal (vertical dimension); adapted from Russell (1980)

Using crowd-sourcing, Mohammad (2018) has produced a lexicon of 20,000 English words, assigning a value to each word for each of the three dimensions[4]. There are, in fact, two equivalent variants of this lexicon: one with values in the range 0 to 1, and another with values in the range -1 to +1. In what follows we use the latter variant. However, it must be stressed that this is an entirely arbitrary decision. When we talk about negative valence and negative arousal we could equally well, if we used the range 0 to 1, talk about low valence and low arousal. As explained in Section 2, we use this lexicon to assign artworks, musical extracts, and our participants' textual responses to points in the VA plane.

---

[4] The lexicon is available at https://saifmohammad.com/WebPages/nrc-vad.html

## 1.4 Research questions

Our first three research questions are concerned with understanding how music affects the appreciation and interpretation of visual art:

> RQ1 Can cross-modal experiences enhance the appreciation of visual art?
>
> RQ2 What connections do people make between the art and the music during the appreciation of the visual art?
>
> RQ3 How is the relationship between art and music affected by the emotions they individually conveyed, and by the congruence and incongruence of those emotions?

Our final research question is practical:

> RQ4 What guidance can be given on the use of music to enhance the experience of visual art?

We review these research questions in Section 4.

# 2 Materials and method

Our approach extends the work of Janzen et al. (2023), who used four pieces of music, representing the four quadrants of the VA plan, but only one artwork. In our work, we chose four artworks, each to represent one of the quadrants of the VA plane. Each participant viewed one of these artworks in association with four pieces of music, again chosen to represent the four quadrants. Subsection 2.1 describes how we chose the artworks, and subsection 2.2 describes how we chose the musical extracts. Subsection 2.3 describes the overall structure of the experiment, including the questions posed to the participants. Section 2.4 describes the questions. Finally, subsection 2.5 describes how we recruited the participants and provides some basic demographic data about them.

## 2.1 Selecting the artworks

In selecting the artworks, we drew on the work of Mohammad and Kiritchenko (2018). They chose 4,105 artworks from WikiArt, selecting in particular from those artworks which WikiArt feature more prominently. They then used crowdsourcing to associate emotion categories with each artwork. They used 19 emotion categories, plus neutral. Each of these emotion categories was represented by an abstract noun, with between one and four additional nouns to help convey the meaning of the emotion. Whilst influenced by the literature on emotions, these categories did not correspond to any established schema. Annotators were asked to tag the image, the title and the image plus title; in each case they were free to use more than one emotion. In choosing our artworks, we used the annotations assigned to the image, ignoring neutral. Thus, from Mohammad and Kiritchenko (2018) we were provided with 4,105 artworks, with for each artwork the proportion of annotators who has tagged the artwork with each of the 19 emotions[5]. We then used the VAD lexicon developed by Mohammad (2018) to calculate the valence and arousal of each of the 19 emotions; in doing this we averaged over the set of words used to describe the emotion. Finally, for each of the 4,105 artworks we took the sum of the valence of each of the emotions, weighted by the proportion of annotators who associated the emotion with the artwork. We ascribed one 'vote' to each annotator; e.g., if an annotator tagged a picture with two emotions, then each emotion's contribution was multiplied by 0.5. We did the same for arousal, thereby locating each artwork at a point in the VA plane. Mirroring the distinction made in subsection 1.1 between induced and perceived emotions in music, Mohammad and Kiritchenko (2018) talk about the various perspectives from which one can label art for emotions, e.g. the emotion the painter is trying to convey or the emotions felt by the individuals in the painting. Their focus, however, was "on the emotions evoked in the observer."

---
[5] This data can be downloaded from https://saifmohammad.com/WebPages/wikiartemotions.html

We wanted to choose four artworks which were most representative of the four quadrants. To do this, we found the four artworks closest, using Manhattan distance, to the extreme points of each quadrant, i.e. (1,1), (1, -1), (-1,1), (-1,-1). In certain cases we rejected the first choice because of concerns about causing distress or offence. In these cases, we simply found another artwork close to the extreme point of the relevant quadrant. As a result, we arrived at the four images shown in Figure 2. The figures were downloaded from WikiArt in jpeg format; all the images used are in the public domain.

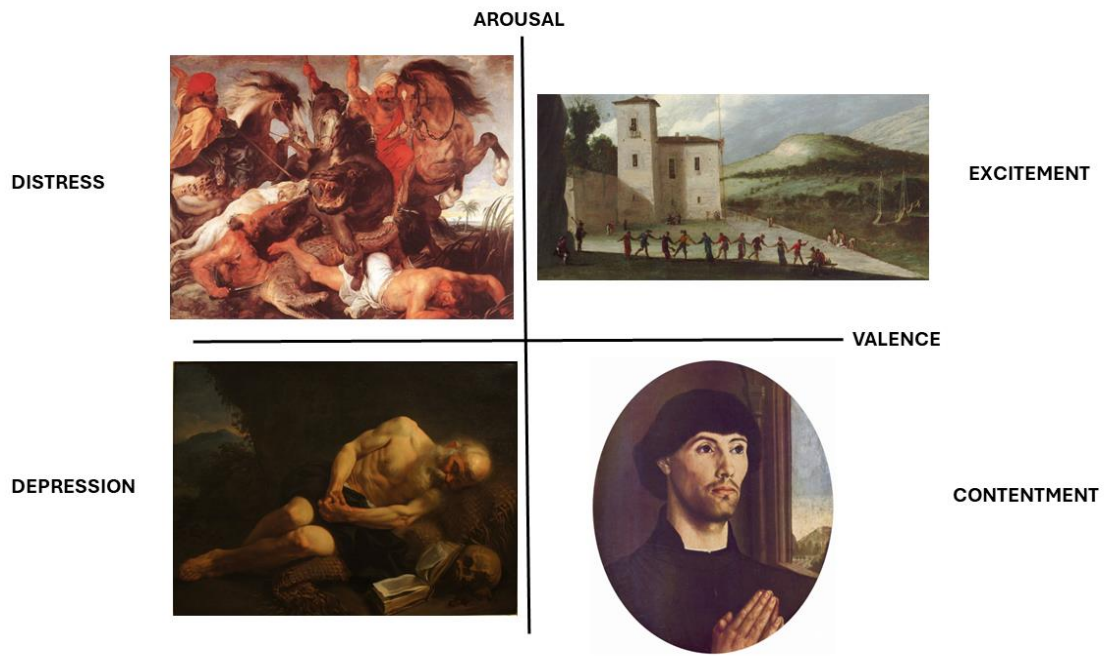

Figure 2. The four images used in the study[6]. They are, working clockwise from top-right: *Festivities on the Coast*, by Agostino Tassi; *Portrait of a Man*, by Hugo van der Goes; *Anachorète endormi*, by Joseph-Marie Vien; and *Hippopotamus and Crocodile Hunt* by Peter Paul Rubens

---

[6] Working from top-right in a clockwise direction, the images are available at:
https://www.wikiart.org/en/agostino-tassi/festivities-on-the-coast-calendimaggio-1620;
https://www.wikiart.org/en/hugo-van-der-goes/portrait-of-a-man-1475;
https://www.wikiart.org/en/joseph-marie-vien/anachor-te-endormi-1751;
https://www.wikiart.org/en/peter-paul-rubens/hippopotamus-and-crocodile-hunt-1616

## 2.2 Selecting the music

As with the artworks, we wanted four pieces of music representing the four quadrants in the VA plane. In fact, we chose two separate groups of four musical extracts. One group was selected in a manner similar to the selection of the artworks. We drew on the work of Aljanaki et al. (2016). They created a database of 400 one-minute musical extracts, all of which were incipits, and divided equally between four genres: classical, rock, pop and electronic. They then used crowd-sourcing to annotate these with the top-level GEMS categories described in subsection 1.2[7]. They changed the names of three of the categories, because they felt that the new names would be better understood[8]. They also provided brief descriptions of each term. We concatenated the name of each emotion category with its description and then, as with the artworks, used the VAD lexicon developed by Mohammad (2018) to calculate a valence and arousal value for each emotion category. Again as with the artworks, we used the annotation data to calculate a valence and arousal value for each of the music extracts, i.e. to locate each extract at a point in the VA plane. Again, we looked for the four extracts which were closest, using Manhattan distance, to the points (1,1), (1,-1),(-1,-1) and (-1,1). We originally intended to restrict our choice to classical extracts, and to avoid lyrics. However, there were no classical extracts in the database at all close to the two negative valence corners. For this reason, we included the whole database  In two cases, we used the second closest extracts because the first closest had lyrics which might be regarded as suggestive.

The other group of musical extracts, all classical, was chosen by one of the authors (NB) to represent the four emotional quadrants, based on her expert knowledge of the classical repertoire. The resulting eight extracts are listed in Table 1. In our implementation, the four extracts chosen algorithmically were embedded in our experiment website. In the table we provide the track ID used in the database; IDs beginning with 2 refer to electronic music, IDs beginning with 3 refer to pop. The four extracts chosen by expert were from YouTube, embedded in our website.

---

[7] The sound files and the annotation data are available at
https://www.projects.science.uu.nl/memotion/emotifydata/
[8] It is worth pointing out here that the original study by Zentner et al. (2008) was carried out in French, and the schema in their paper and on the website is a translation.

Table 1. The eight musical extracts. For the extracts chosen by algorithm, numbers in brackets indicate the ID of the extract on the source database. For the extracts chosen by expert, we provide the URLs as footnotes.

N.B.   † contains lyrics;

‡ contains vocals, but with "an almost wordless text" (Swayne, 2024)

| | algorithmically-chosen music | |
|---|---|---|
| | negative valence | positive valence |
| positive arousal | *Ping Heng* - Solace<br>from: Balance<br>(262) | *Sunset*† - D J Markitos<br>from: Evolution of the Mind<br>(228) |
| negative arousal | *Below*† – Lisa DeBenedictis<br>from: Tigers<br>(370) | *Rubber* – Williamson<br>from: A few things to hear before we all blow up (354) |
| | expertly-chosen music | |
| | negative valence | positive valence |
| positive arousal | *Threnody for the Victims of Hiroshima*[9] – Penderecki | *In The Upper Room: Dance No.9*[10] - Glass |
| negative arousal | *Cry, Op. 27: I. Void - Light – Darkness*‡[11] - Swayne | *Spiegel im Spiegel*[12] - Arvo Pärt |

## 2.3 The experiment

The experiment was hosted on an Open University server. As will be explained in detail in the next subsection, there were two types of questions: multiple-choice questions; and open-ended questions giving participants an opportunity to provide a more detailed response. Initially, 18 experimental sessions were carried out by the experimenter over Microsoft Teams. In these cases, participants were offered the choice between responding orally to the detailed questions or typing their responses. For the 16 participants who responded orally, a transcript was obtained from Microsoft Teams. An additional 120 participants took part without the intervention of the experimenter and were required to type their responses. Each participant saw one of four artworks, and listened to one of the two groups of musical extracts. Consequently, there were eight variants in all. Table 2 indicates the number of participants assigned to each variant. In the table we adopt the convention, which we use throughout this paper, of indicating the quadrant in the VA plane in the form, e.g. (P/P) or (P/N). Here, the first letter indicates the sign of the valence, i.e. either positive or negative; whilst the second letter similarly indicates the sign of the arousal.

---

[9] https://www.youtube.com/watch?v=Pu371CDZ0ws&t=225s

[10] https://www.youtube.com/watch?v=wPdLu8GQprE&t=180s

[11] https://www.youtube.com/watch?v=iuo8cDf304Q&t=120s

[12] https://www.youtube.com/watch?v=JxYZu9piR64&t=30s

Table 2. Number of participants

|  | music chosen by algorithm | music chosen by expert | total for artwork |
|---|---|---|---|
| Festivities on the Coast (P/P) | 15 | 16 | 31 |
| Portrait of a Man (P/N) | 19 | 20 | 39 |
| Anachorète endormi (N/N) | 16 | 20 | 36 |
| Hippopotamus and Crocodile Hunt (N/P) | 16 | 16 | 32 |
| Total | 66 | 72 | 138 |

## 2.4 The Questions

The questions were divided into four sections. An initial section asked for a limited amount of information about the participants. The second section sought the participant's reaction to the artwork, before listening to music. The third section sought the participant's reaction to the artwork after listening to a musical extract. Participants heard four pieces of music and were required to complete this section after each. The order in which the musical extracts were heard was randomised to compensate for any order effects. Finally, a fourth section contained some questions regarding the whole experience.

The study was approved by The Open University Human Research Ethics Committee, reference number 2023-0045-3. Before taking part in the study, participants signed a consent form which gave permission for the information provided to be used in the study and published in anonymised form. This included permission to be quoted anonymously. The 18 participants who took part in the experiment carried out by the experimenter over Microsoft Teams, also gave permission that the session be recorded.

### 2.4.1 Participant information

The first section contained four questions. Two of these were demographic. Participants were asked their age, in ranges of decades. They were also asked their gender. For both these questions, there was also a 'prefer not to say' option. The final two questions asked how frequently the participant looked at art and how much time they spent listening to music:
- How often do you look at art? This might be by visiting a gallery, looking at art online, watching a TV programme about art, or looking at a book about art.
- On average, how much time each day do you spend listening to music? This might be listening to live music, recorded music, music streaming services or radio.

The possible responses to these questions, plus the breakdown of responses between the participants are shown in subsection 2.5.

### 2.4.2 Initial viewing of artwork

The next section, to obtain participants' impression of the artwork before listening to music, contained four questions. The first two of these required oral or typed responses:
- What do you notice about the artwork? What are you thinking about the artwork?
- How does the artwork make you feel?

Participants did not always distinguish between these two questions, i.e. between 'noticing' and 'thinking', and 'feeling'; e.g. sometimes answering the second question as part of the first.

Consequently, for the sentiment analysis in subsection 3.3, responses to these two questions were considered together.

The other two questions were multiple choice questions with one permitted response:
- How meaningful do you find the artwork?
- How pleasant do you find the artwork?

Table 3 shows the possible responses to the first of these questions. Responses to the second followed the same pattern, i.e. 'not at all pleasant' … 'extremely pleasant'.

Table 3. Meaningfulness question

| How meaningful do you find the artwork? |
|---|
| extremely meaningful |
| very meaningful |
| somewhat meaningful |
| not so meaningful |
| not at all meaningful |

These questions reflect the cognitive and affective aspects of the aesthetic experience, as described by Leder and Nadal (2014).

### 2.4.3 After listening to each piece of music

After listening to each piece of music, there were initially two open-ended questions:
- Did you now notice anything else about the artwork? Were you drawn to different things about the artwork?
- How does the artwork make you feel with this piece of music? Is your feeling different from when you viewed the artwork without music?

As with the first two questions mentioned in subsection 2.4.2, responses to these two questions needed to be considered together.

Participants were then asked to indicate what connections they were making between the artwork and the music. The question and the possible responses are shown in Table 4. As can be seen, the first four of these responses correspond to the modes of connection discussed towards the end of subsection 1.1 above. Participants were free to tick as many boxes as they liked, or none. It became clear during the initial 18 supervised experiments, that the latter option needed to be made explicit. Therefore, for the remaining 120 unsupervised experiments participants were able to indicate that they found no connections. In the analysis this option was ignored; it was simply there to make absolutely clear to participants that they did not need to specify any mode of connection. These questions can be seen as probing the range of psychological processes involved in the aesthetic experience, as described by Leder and Nadal (2014).

Table 4. Question about mode of connection between artwork and music
Letters in brackets indicates the code which is used subsequently, e.g. in Figure 6.

| **Which of the following connections were you making between the artwork and the music? Tick any which apply.** |
| --- |
| A theme of idea that connected the music or artwork. (T) |
| Something in the artwork moving in time to the music. (M) |
| A story that connected the music and the artwork. (S) |
| An emotional connection between the music and the artwork. (E) |
| Other type of connection. (O) |

There was then an open-ended question:
- Please explain your previous answer.

Finally, there were three multiple choice questions with a single permitted response:
- How connected do you find the artwork and the music?
- How meaningful do you find the artwork with this piece of music?
- How pleasant do you find the artwork with this piece of music?

The response options for the second and third of these questions are as described in subsection 2.4.2. The response options for the first question follows the same pattern, i.e. ranging from 'not all connected' to 'extremely connected'.

### 2.4.4  Final section

The final section contained three questions. The first question was a multiple choice question with a single permitted response:
- Overall, how helpful were the different pieces of music in developing an appreciation of the artwork?

The permitted responses were on a five-point scale following the same pattern as the other single response questions, i.e. ranging from 'not at all helpful' to 'extremely helpful.

The last two questions were open-ended:
- Please explain your previous answer.
- Do you have any other comments about the experience?

### 2.5  The participants

The majority of participants were recruited through an Open University *Microsoft Viva Engage*[13] platform. This platform is used by employees of the Open University, but not students. Other participants were recruited through an internal university mailing list, a collaborative project mailing list[14], and personal invitation.

Tables 5 and 6 give the age distribution, and the distribution by gender, for each of the two groups and overall. Table 7 shows the frequency of looking at art amongst the participants, and Table 8 shows the distribution of time per day spent listening to music.

---
[13] Formerly known as *Yammer*.
[14] Polifonia, see https://polifonia-project.eu/

Table 5. Age distribution of participants

|  | music chosen by algorithm | music chosen by expert | overall |
|---|---|---|---|
| under 20 | 0 | 0 | 0 |
| 20 to 29 | 5 | 1 | 6 |
| 30 to 39 | 12 | 7 | 19 |
| 40 to 49 | 20 | 18 | 38 |
| 50 to 59 | 19 | 24 | 43 |
| 60 to 69 | 7 | 17 | 24 |
| 70 or over | 2 | 5 | 7 |
| prefer not to say | 1 | 0 | 1 |

Table 6. Gender distribution of participants

|  | music chosen by algorithm | music chosen by expert | overall |
|---|---|---|---|
| female | 50 | 51 | 101 |
| male | 13 | 20 | 33 |
| other | 2 | 0 | 2 |
| prefer not to say | 1 | 1 | 2 |

Table 7. Frequency of viewing art

|  | music chosen by algorithm | music chosen by expert | overall |
|---|---|---|---|
| never | 0 | 0 | 0 |
| less than once a year | 5 | 4 | 9 |
| at least once a year, but not every month | 19 | 17 | 36 |
| at least once a month, but not every week | 16 | 22 | 38 |
| at least once a week, but not every day | 16 | 15 | 31 |
| every day | 10 | 14 | 24 |

Table 8. Time spent listening to music, per day

|  | music chosen by algorithm | music chosen by expert | overall |
|---|---|---|---|
| no time at all | 0 | 0 | 0 |
| less than half an hour | 10 | 8 | 18 |
| between half an one hour | 16 | 17 | 33 |
| between one and two hours | 16 | 19 | 35 |
| between two and three hours | 11 | 10 | 21 |
| more than three hours | 13 | 18 | 31 |

# 3  Results

In subsection 3.1 we investigate quantitatively participants' responses to multiple choice questions. In subsection 3.2 we discuss participants' responses to the open-ended questions. Finally, in subsection 3.3 we apply sentiment analysis to these responses to determine what additional insight this provides.

## 3.1  Multiple choice questions

We begin, in subsection 3.1.1, by investigating how participants perceived the connectedness between artwork and music. In subsections 3.1.2 and 3.1.3 we discuss the results of our questions about meaningfulness and pleasantness. We have already noted that Russell (2003) had found a lack of significant correlation between the two, and in our study we made meaningfulness and pleasantness the subject of separate questions. Then, in subsection 3.1.4, we investigate correlations between connectedness, meaningfulness and pleasantness in our results. In subsection 3.1.5 we investigate how helpful participants found the music in developing an appreciation of the artwork. Finally, in subsection 3.1.6 we provide a brief summary.

### 3.1.1  Connectedness

As part of our first research question, we asked whether proximity in the VA plane would enhance the aesthetic experience of the artwork. We hypothesized that such an enhancement would be dependent on a close emotional connection between the artwork and the music. On the basis of the emotional mediation theory, we expected that this close connection would arise where artwork and music were in the same quadrant of the VA plane. To test this, we examined the responses to the question 'How connected do you find the artwork and the music?' which we asked after participants viewed an artwork with each piece of music. We conducted a Friedman rank sum test with four conditions, depending on whether the artwork and music were:
- in the same quadrant;
- in opposing quadrants, i.e. with differently signed valence and arousal;
- in different quadrants, with the same signed valence and differently signed arousal;
- in different quadrants, with differently signed valence and the same signed arousal.

Figure 3 shows the median and quartiles for the four conditions, overall and separately for the music chosen by algorithm and the music chosen by expert. Overall, there was no significant difference between the conditions ($\chi^2(3, N = 552) = 1.5042$, $p = 0.6813$). Similarly, there was no significant difference between the conditions for music chosen by algorithm ($\chi^2(3, N = 264) = 2.0225$, $p = 0.5678$); nor for the music chosen by expert ($\chi^2(3, N = 288) = 2.0594$, $p = 0.5602$).

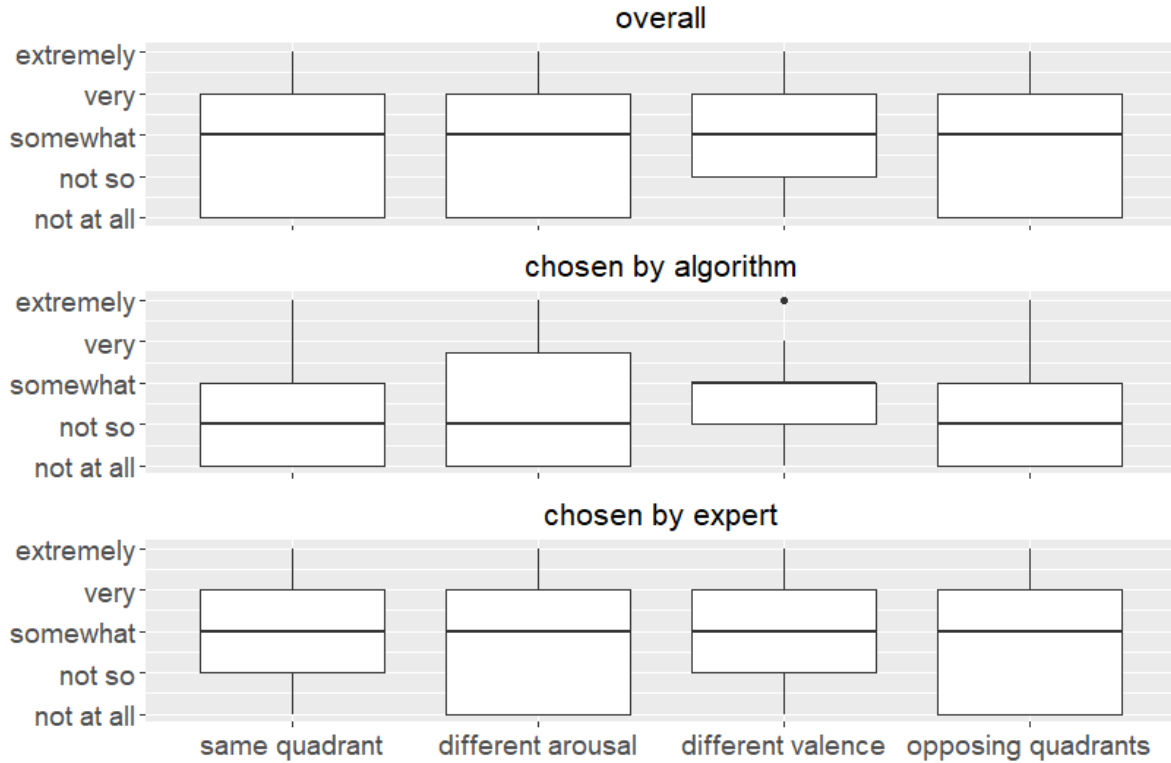

Figure 3 Connectedness responses by alignment

When we consider not the alignment between artwork and music, but rather the valence and arousal of the music, we do see significant differences in connectedness. Figure 4 shows the mean and quartile connectedness by quadrant occupied by music in VA plane. A Friedman rank sum test, taking the conditions as being the quadrant in the VA plane occupied by the music showed a significant difference ($\chi^2(3, N = 552) = 28.06$, $p = 3.529e-06$). The same was true separately for the groups hearing the music chosen by algorithm ($\chi^2(3, N = 264) = 48.157$, $p = 1.972e-10$) and by expert ($\chi^2(3, N = 288) = 9.1058$, $p = 0.02792$) music. This suggests that certain pieces of music foster connectivity more than others. However, these Friedman rank sum tests are aggregating effects over all artworks. There are considerable differences over the various music – artwork combinations. This can be seen in Table 9, which shows the results of Friedman tests for each artwork separately

Table 10 shows, for the music chosen by algorithm, the result of pairwise Wilcoxon tests for the four conditions shown in Figure 4. Taking the table and figure together, it can be seen that both musical extracts with negative valence were significantly more connected to the artwork than each of the musical extracts with positive valence (see top two rows of table), and that positive arousal significantly amplified the effect of negative valence (see bottom row). Table 11 shows the results of pairwise Wilcoxon tests for the music chosen by expert, revealing a different picture. The music with positive valence and negative arousal is significantly more connected than the two music extracts with positive arousal. Apart from that, there are no significant differences. Of course, these comments are generalized over the four artworks, and there are variations for individual artworks.

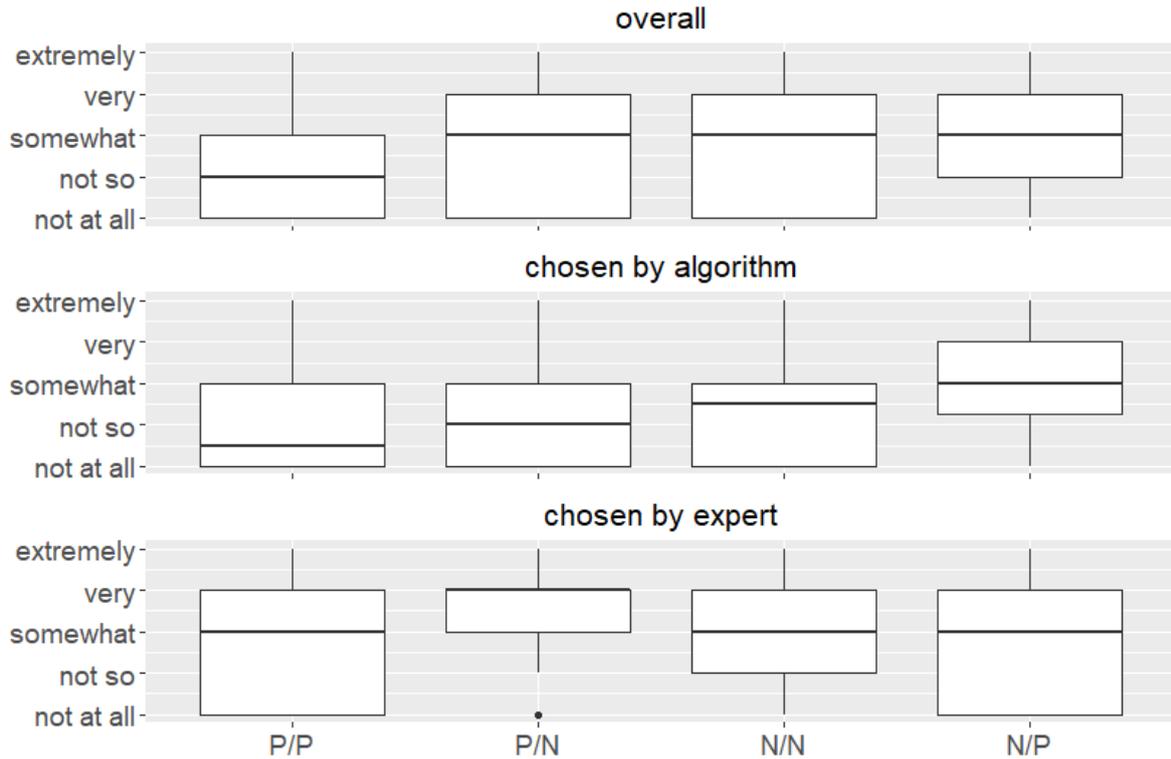

Figure 4 Connectedness responses by music quadrant
For the horizontal axis labels, the first letter refers to the sign of the valence, the second to the sign of the arousal, e.g. P/P refers to music with positive valence and positive arousal

Table 9. Results of Friedman tests for variation in connectedness across the musical extracts, by artwork[15]

|  | music chosen by algorithm | music chosen by expert |
|---|---|---|
| Festivities on the Coast (P/P) | $\chi^2(3, N = 60) = 9.528$, $p = 0.02304$ | $\chi^2(3, N = 64) = 8.1176$, $p = 0.04364$ |
| Portrait of a man (P/N) | $\chi^2(3, N = 76) = 21.361$, $p = 8.859e-05$ | $\chi^2(3, N = 80) = 6.6708$, $p = 0.08316$ |
| Anachorète endormi (N/N) | $\chi^2(3, N = 64) = 10.897$, $p = 0.01229$ | $\chi^2(3, N = 80) = 9.6043$, $p = 0.02225$ |
| The Hippopotamus and Crocodile Hunt (N/P) | $\chi^2(3, N = 64) = 15.537$, $p = 0.001411$ | $\chi^2(3, N = 64) = 2.625$, $p = 0.4531$ |

---

[15] No correction for multiple statistical tests has been applied. To apply a Bonferroni correction, the usual threshold of p = 0.05 should be reduced to p = 0.05/8, i.e. 0.00625.

Table 10 Pairwise Wilcoxon tests of difference in connectivity for musical extracts chosen by algorithm, over all artworks
* p < 0.05; ** p < 0.01; *** p < 0.001

|  | P/N | N/N | N/P |
|---|---|---|---|
| P/P | V = 455.5<br>p = 0.6416 | V = 290<br>p = 0.005322 *** | V = 211<br>p = 4.367e-07 *** |
| P/N |  | V = 375<br>p = 0.04233 * | V = 156<br>p = 1.531e-06 *** |
| N/N |  |  | V = 768.5<br>p = 0.0003048 *** |

Table 11 Pairwise Wilcoxon tests of difference in connectivity for musical extracts chosen by expert, over all artworks
* p < 0.05; ** p < 0.01; *** p < 0.001

|  | P/N | N/N | N/P |
|---|---|---|---|
| P/P | V = 372.5<br>p = 0.009812 ** | V = 510.5<br>p = 0.2113 | V = 522.5<br>p = 0.4982 |
| P/N |  | V = 859.5<br>p = 0.1161 | V = 732.5<br>p = 0.0332 * |
| N/N |  |  | V = 543<br>p = 0.6423 |

Kruskal-Wallis tests of the effect of artwork show no significant differences, neither for both music groups together ($\chi^2(3)$ = 7.4497, p = 0.05887), nor for the music chosen by algorithm ($\chi^2(3)$ = 4.4489, p = 0.2169), nor for the music chosen by expert ($\chi^2(3)$ = 3.8107, p = 0.2826). Comparison with the analogous treatment of music given above suggests that music is the dominant cause of variability in connectedness. Indeed, for the music chosen by algorithm the median connectedness was 'not so connected' for all artworks, except for the artwork with negative valence and positive arousal, where the median connectedness was 'somewhat connected'. For the music chosen by expert, the median connectedness was 'somewhat connected' for all artworks.

Figure 5 provides a detailed picture, showing boxplots for each combination. We can make some immediate comments. When we look horizontally and vertically across the bar-charts, we frequently see a considerable variation in responses for each artwork, and for each musical extract. A good example of this is music with positive valence and arousal, in both groups of music. On the other hand, there are cases where there is more consistency. For example, if we consider the bottom-row for the expertly-chosen music, we see that the artwork with negative valence and positive arousal (*Hippopotamus and Crocodile Hunt*, by Rubens) connected well with all four pieces of music. Additionally, if we consider the second column from the left for the music chosen by algorithm, we see that the music with positive valence and negative arousal (*Rubber*, by Williamson) had little connection with any of the artworks.

If, for each music group, we consider the diagonal from top-left to bottom-right, we see the four situations in which artwork and music are in the same quadrant. It is apparent that there is an appreciable difference between the four distributions of responses, again refuting our original hypothesis that in these situations there would be considerable connectedness. It is also interesting to compare corresponding boxplots in the two groups of music, where we frequently see appreciable differences in the distributions of responses; a good example of this is illustrated by the two top-right boxplots, representing music with negative valence and positive arousal and artworks with positive valence and positive arousal.

For each artwork, we see considerable variation in connectedness between the four pieces of music chosen by algorithm. For the music chosen by expert, this is not always the case. As illustrated in Table 9, for the music chosen by algorithm and for each artwork, Friedman tests revealed significant differences in degree of connectiveness across each of the musical extracts. For the music chosen by expert, there were significant differences across the musical extracts for the artworks with positive valence and arousal (*Festivities on the Coast*), and for the artworks with negative valence and arousal (*Anachorète endormi*). This was not the case for the artwork with negative valence and positive arousal (*Hippopotamus and Crocodile Hunt*) nor for the artwork with positive valence and negative arousal (*Portrait of a Man*).

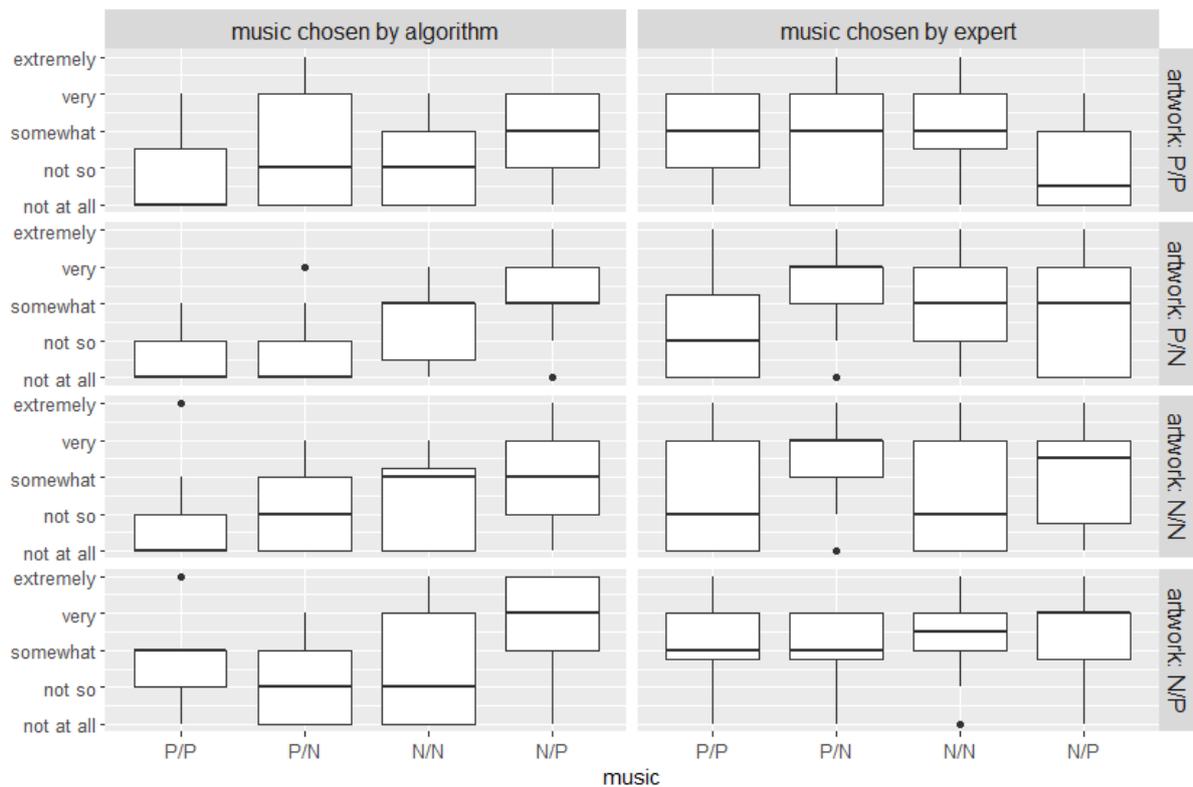

Figure 5. Connectedness responses for each artwork - music combination[16]

Finally, there was a significant difference between participants who heard the music chosen by algorithm and those who heard the music chosen by expert (Mann-Whitney test: W = 46798, p = 1.354e-06). For the former, the median was 'not so connected'; for the latter the median was 'somewhat connected'. With a few exceptions, this distinction can be seen at a detailed level in Figure 5.

The preceding discussion suggests that emotion is far from the only determinant of connectivity between artwork and music. As we noted in subsection 2.4.3, after each piece of music we asked participants what connections they were making between the artwork and the music, from the options shown in Table 4. Figure 6 shows the percentage of participants indicating each option, for both groups of music. The bottom right-hand plot shows the distribution of responses overall. For both sets of music, *emotion* was the most popular response, with *story* and *theme* also quite popular. *Movement* was less popular, although chosen by appreciably more of those who heard the music chosen by expert; whereas *other* was chosen by more of those who heard the music chosen by algorithm. In interpreting this figure, it should be borne in mind that participants were free to indicate more than one type of connection. On the other hand, on 29% of the cases no connection was indicated[17].

---

[16] In the boxplot, thick central line represents median, extremities of box represent 25th and 75th percentiles, the upper vertical lines ('whiskers') extend to the largest value no further than 1.5 x inter-quartile range from the upper quartile, the lower vertical lines extend to the smallest value no further than 1.5 x interquartile range from the lower quartile, and any points beyond these two limits are plotted individually, see https://ggplot2.tidyverse.org/reference/geom_boxplot.html

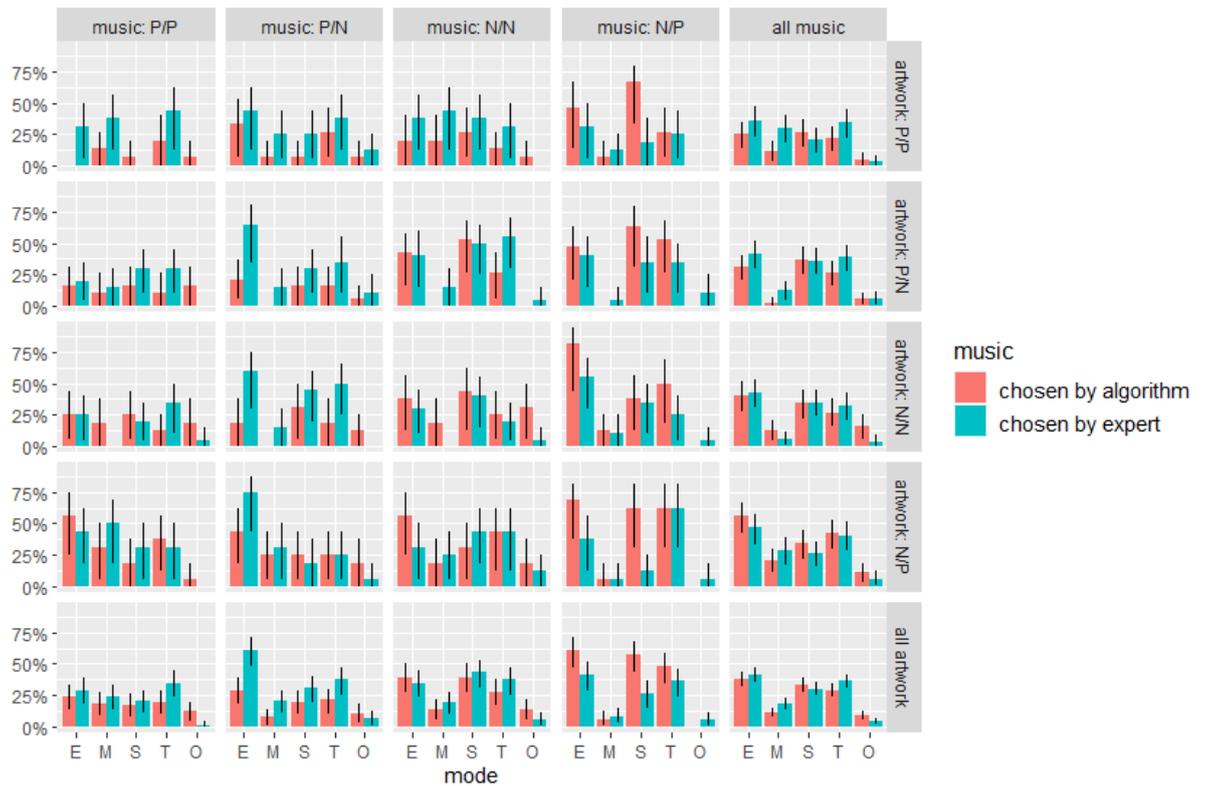

Figure 6. Connections between artwork and music
Horizontal axis shows modes of connection indicated by participants; codes are shown in Table 4; percentages add up to more than 100% because multiple responses were allowed.

### 3.1.2 Meaningfulness

Figure 7 shows, the effect of the music on the perceived meaningfulness. The figure shows an overall reduction in meaningfulness for both groups, although this is less the case for the music chosen by expert. Moreover, a Mann-Whitney test revealed that the music chosen by algorithm caused a significant decrease in meaningfulness (W = 11595, p-value = 1.656e-05). However, the music chosen by expert had no significant effect on meaningfulness (W = 11312, p-value = 0.2185)[18]. Consistent with this, a Mann-Whitney test showed a significant difference between the meaningfulness after listening to the two groups of music (W = 44169, p = 0.0007243).

---

[17] I.e. 159 out of 552 (138 participants x 4 pieces of music).
[18] In both these cases, because for each participant there was one meaningfulness value without music and four values with music, we conducted an unpaired test.

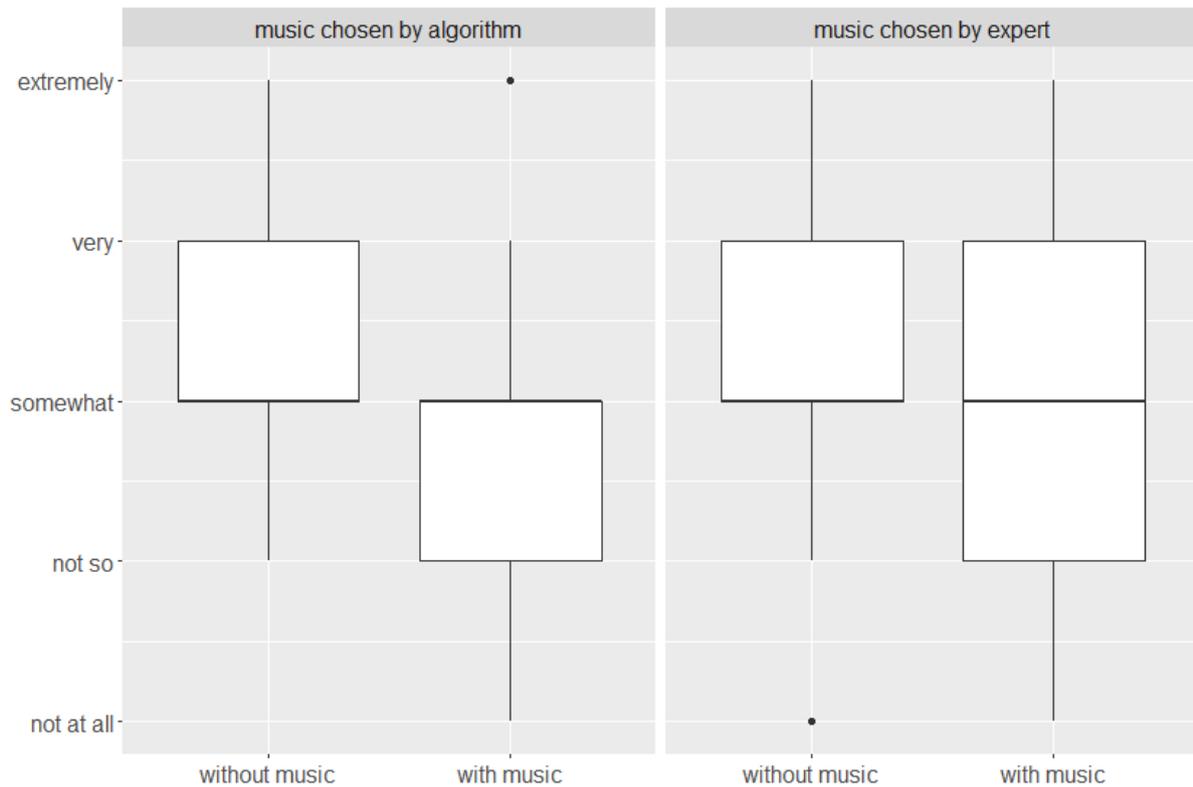

Figure 7. Effect of music on meaningfulness of artwork

Figure 8 presents a more detailed picture by showing the results of the meaningfulness question for each artwork-music combination, and also for the initial response before any music was heard. We consider, first, the response before any music was heard. Combining the two groups of participants, a Kruskal-Wallis test reveals a significant difference across the artworks ($\chi^2(3, N = 138) = 8.4084$, $p = 0.03828$). From the first column of Figure 8, for each of the two groups of participants, we can see that the artwork with negative valence and arousal (*Anachorète endormi*) had the highest meaningfulness; for both groups it had a median meaningfulness between 'somewhat' and 'very'. Although not apparent from Figure 8, when combining data from both groups, the artwork with positive valence and arousal (*Festivities on the Coast*) had least meaningfulness. This may well reflect the serious nature of *Anachorète endormi*. Of course, meaningfulness is subjective. One can imagine that *Festivities on the Coast* might have meaningfulness to someone with particular memories of similar holiday activities.

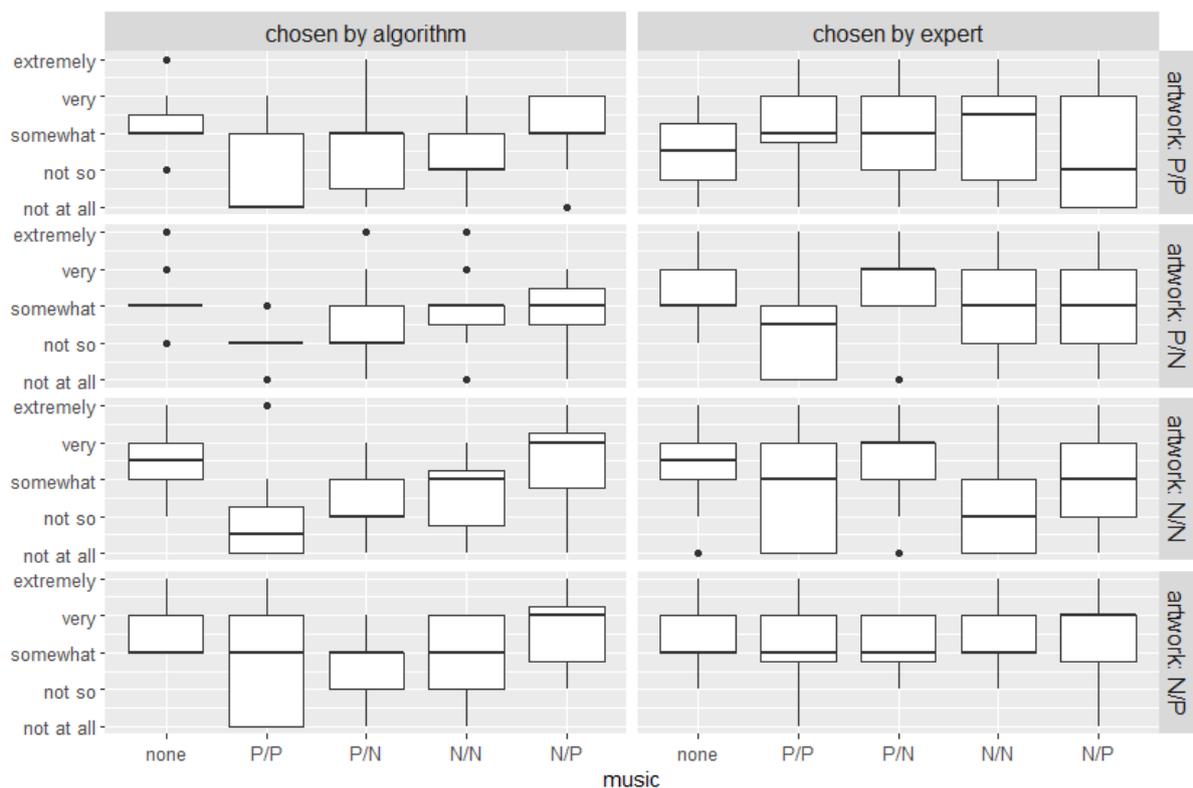

Figure 8. Meaningfulness responses without music, and for each artwork – music combination.

Table 12 shows the p-values resulting from a Wilcoxon test comparing meaningfulness of each artwork before listening to music, to meaningfulness after listening to each piece of artwork. The table enables a number of comments. Three of the pieces of music had no significant effect on the meaningfulness of any of the artworks. These were the two pieces of music with negative valence and positive arousal, and the piece chosen by expert with positive valence and negative arousal. On the other hand, the piece chosen by algorithm with positive valence and positive arousal had a significant effect on the meaningfulness of each of the artworks. For the other four pieces there was a more varied pattern. Comparison with Figure 8 shows that, wherever there was a significant effect, it was to diminish the meaningfulness of the artwork. There are cases where Figure 8 suggests that a piece of music increased the meaningfulness of an artwork, but not significantly. As shown in Table 12, the sample size for each of these Wilcoxon tests was quite small, and it may be that, with a larger sample size there would be evidence of a significant, albeit small, increase in meaningfulness. A particularly interesting case, where Figure 8 shows an increase in meaningfulness can be seen in the bottom right corner plot for each of the two music groups. Here, the artwork had negative valence and positive arousal (*Hippopotamus and Crocodile Hunt*) whilst both pieces of music similarly had negative valence and positive arousal (*Ping Heng* and *Threnody for the Victims of Hiroshima*). This does suggest a resonance between the music and the artwork, determined by similar valence and arousal. Two other cases which suggest a similar resonance arise from the two pieces of music chosen by expert with positive valence, when played with artworks in the same quadrant of the VA plane. However, considering the relevant diagonals in Figure 8, we can see that this resonance effect does not always occur when artwork and music are in the same quadrant.

We can also see that, overall, the music chosen by expert had less effect than that chosen by algorithm; for the former there was only a significant effect with three artwork – music combinations; for the latter with ten combinations. This is consistent with the overall picture we reported at the beginning of this subsection.

Table 12. Results of a Wilcoxon test, showing significance of effect of music on meaningfulness[19]
* p < 0.05; ** p < 0.01

|  | | music chosen by algorithm | | | |
|---|---|---|---|---|---|
|  | N | P/P | P/N | N/N | N/P |
| artwork P/P | 15 | V = 75.5<br>p = 0.0037 ** | V = 23.5<br>p = 0.1241 | V = 76<br>p = 0.0289 * | V = 17<br>p = 0.5328 |
| artwork P/N | 19 | V = 78<br>p = 0.0021 ** | V = 93<br>p = 0.0078 ** | V = 39<br>p = 0.243 | V = 42<br>p = 0.4359 |
| artwork N/N | 16 | V = 112<br>p = 0.0030 ** | V = 70<br>p = 0.0150 * | V = 65<br>p = 0.0404 * | V = 45.5<br>p = 1 |
| artwork N/P | 16 | V = 57<br>p = 0.0341 * | V = 72<br>p = 0.0096 ** | V = 85<br>p = 0.0406 * | V = 47.5<br>p = 0.914 |
|  | | music chosen by expert | | | |
|  | N | P/P | P/N | N/N | N/P |
| artwork P/P | 16 | V = 12<br>p = 0.1121 | V = 29<br>p = 0.2479 | V = 15<br>p = 0.214 | V = 38.5<br>p = 0.6493 |
| artwork P/N | 20 | V = 109<br>p = 0.0049 ** | V = 38.5<br>p = 0.3693 | V = 58.5<br>p = 0.3692 | V = 63<br>p = 0.2238 |
| artwork N/N | 20 | V = 71<br>p = 0.0121 * | V = 25<br>p = 0.8343 | V = 85<br>p = 0.0055 ** | V = 69<br>p = 0.1007 |
| artwork N/P | 16 | V = 25.5<br>p = 0.3046 | V = 30.5<br>p = 0.7912 | V = 10<br>p = 0.5393 | V = 28.5<br>p = 0.4012 |

### 3.1.3 Pleasantness

We analyse pleasantness in the same way as meaningfulness. Figure 9 provides a top-level picture. Overall, both groups of music depressed the perceived pleasantness of the artwork. A Mann-Whitney test revealed that the both the music chosen by algorithm (W = 10310, p-value = 0.01701) and the music chosen by expert (W = 12078, p-value = 0.02531) caused a significant decrease in pleasantness. A Mann-Whitney test showed no significant difference in pleasantness after listening to the two groups of music (W = 38465, p = 0.8046).

---

[19] As before, no correction for multiple tests has been applied. To apply a Bonferroni correction, the usual threshold of p = 0.05 should be reduced to p = 0.05/32, i.e. 0.0015625.

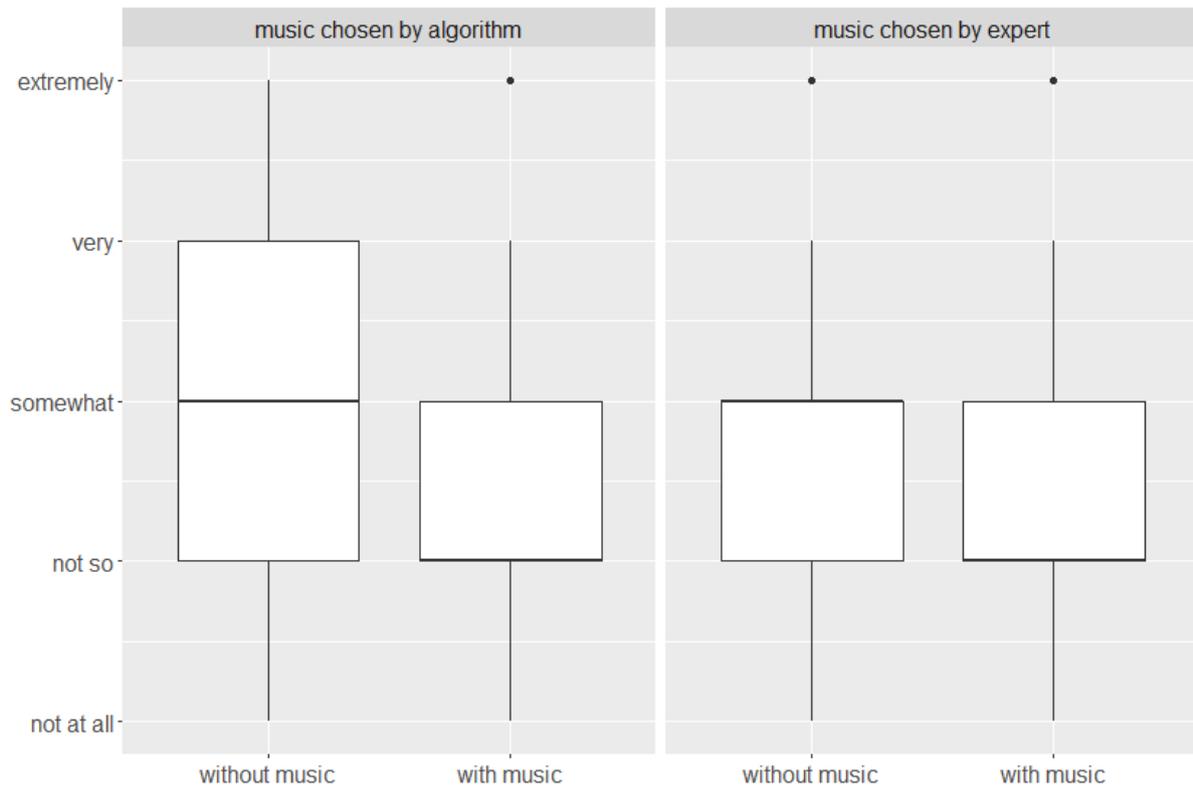

Figure 9. Effect of music on pleasantness of artwork

Figure 10 shows the results of the meaningfulness question for each artwork-music combination, and also for the initial response before any music was heard. As in the last subsection, we start by considering the response before any music was heard. Combining the two groups of participants, a Kruskal-Wallis test reveals a significant difference across the artworks ($\chi^2$(3, N = 138) = 38.301, p = 2.441e-08). The artwork with positive valence and arousal (*Festivities on the Coast*) was judged the most pleasant, followed by the artwork with positive valence and negative arousal (*Portrait of a Man*), then the artwork with negative valence and negative arousal (*Anachorète endormi*), and finally the artwork with negative valence and positive arousal (*Hippopotamus and Crocodile Hunt*). This pattern can be seen in the leftmost columns for both of the two groups in Figure 10. Given that valence is an equivalent term for pleasantness, the preference for positively valenced artworks serves to confirm that our participants were in agreement with the original artwork annotators. The fact that the most pleasant and least pleasant artworks had positive arousal suggests that arousal may have an amplifying effect, i.e. increasing the pleasantness of positively valenced and decreasing the pleasantness of negatively valenced.

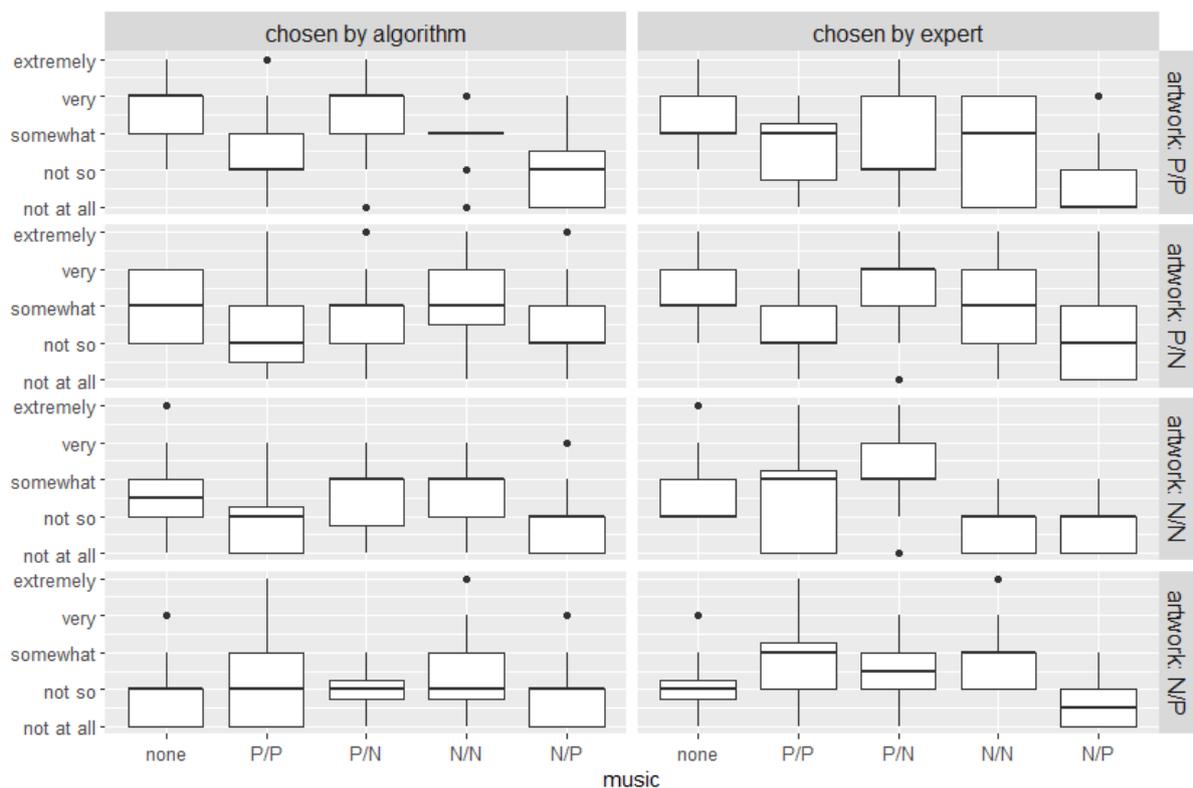

Figure 10. Pleasantness responses without music, and for each artwork – music combination.

Table 13 shows the p-values resulting from a Wilcoxon test comparing pleasantness of each artwork before listening to music, to pleasantness after listening to each piece of artwork. There were two pieces of music which had no significant effect on the pleasantness of any artwork; these were the two pieces with negative arousal chosen by algorithm. All the other pieces had a more varied effect on the artworks. For the music chosen by algorithm, there were five artwork – music combinations where the effect of the music was significant. In each of these cases, the music reduced the pleasantness of the artwork. For the music chosen by expert, there were eight artwork – music combinations where the effect of the music was significant. For six of these the effect was negative. However, for the artwork with negative valence and positive arousal (*Hippopotamus and Crocodile Hunt*), the two pieces of music with negative arousal (*Cry, Op. 27: I. Void - Light – Darkness* by Swayne and *Spiegel im Spiegel* by Pärt) had the effect of increasing pleasantness. In these cases we are comparing with a low base, since this artwork was regarded as least pleasant when viewed without music. Similarly, it is perhaps not surprising that no music was able to increase the pleasantness of the artwork with positive valence and arousal (*Festivities on the Coast*), since this artwork was regarded as most pleasant when viewed without music.

Finally, considering the relevant diagonals in the two music groups apparently disproves the idea that an increase in pleasantness is likely to occur when artwork and valence are in the same quadrant of the VA plane.

Table 13. Results of a Wilcoxon test, showing significance of effect of music on pleasantness[20]. Statistics in green and underlined indicate where a significant effect was positive.
\* p < 0.05; \*\* p < 0.01

|  | | music chosen by algorithm | | | |
|---|---|---|---|---|---|
|  | N | P/P | P/N | N/N | N/P |
| artwork P/P | 15 | V = 75<br>p = 0.0044 ** | V = 34<br>p = 0.5242 | V = 59<br>p = 0.1164 | V = 84<br>p = 0.0072 ** |
| artwork P/N | 19 | V = 82<br>p = 0.00968 ** | V = 32.5<br>p = 0.627 | V = 47<br>p = 0.7456 | V = 78<br>p = 0.1038 |
| artwork N/N | 16 | V = 72.5<br>0.0056 ** | V = 43.5<br>p = 0.3521 | V = 18<br>p = 0.6078 | V = 61.5<br>p = 0.0093 ** |
| artwork N/P | 16 | V = 21.5<br>p = 0.3117 | V = 29.5<br>p = 0.7805 | V = 32.5<br>p = 0.1866 | V = 36<br>p = 0.8128 |
|  | | music chosen by expert | | | |
|  | N | P/P | P/N | N/N | N/P |
| artwork P/P | 16 | V = 44.5<br>p = 0.08624 | V = 47<br>p = 0.2091 | V = 64.5<br>p = 0.0422 * | V = 78<br>p = 0.0022 ** |
| artwork P/N | 20 | V = 109<br>p = 0.0043 ** | V = 28<br>p = 0.0963 | V = 46<br>p = 0.2362 | V = 142<br>p = 0.0016 ** |
| artwork N/N | 20 | V = 79<br>p = 0.9225 | V = 26.5<br>p = 0.0541 | V = 100.5<br>p = 0.0157 * | V = 78<br>p = 0.0019 ** |
| artwork N/P | 16 | V = 16<br>p = 0.0713 | <u>**V = 8**</u><br><u>**p = 0.0437 \***</u> | <u>**V = 4**</u><br><u>**p = 0.0148 \***</u> | V = 65<br>p = 0.1672 |

### 3.1.4 Correlations

Table 14 shows the Spearman correlation coefficients between reported meaningfulness and pleasantness, overall and for the two music groups; and in each case, initially without music and then with music. In the initial case without music the coefficients are low, confirming the findings of Russell (2003); although our correlations are all positive, whereas Russell's (2003) correlations ranged from -0.34 to +0.01. However, meaningfulness and pleasantness were more highly correlated after listening to music. It is not entirely clear what is causing this. It may be that the effect of the music is to alter the reported meaningfulness and pleasantness in the same direction, frequently in the negative direction, and hence this causes greater correlation.

---

[20] As with Table 12, applying a Bonferroni correction would reduce p = 0.05 to p = 0.0015625

Table 14. Spearman's rank correlation between reported meaningfulness and pleasantness;
* p < 0.05; ** p < 0.01; *** p < 0.001

|  | without music | with music |
|---|---|---|
| music chosen by algorithm | 0.07<br>p = 0.5546 | 0.38<br>p = 2.345e-10 *** |
| music chosen by expert | 0.24<br>p = 0.0457 * | 0.58<br>p < 2.2e-16 *** |
| overall | 0.17<br>p = 0.0469 * | 0.49<br>p < 2.2e-16 *** |

Table 15 shows the correlation between reported connectedness and meaningfulness, and between connectedness and pleasantness. It can be seen that connectedness correlates much more with meaningfulness than it does with pleasantness.

Table 15. Spearman's rank correlation between reported connectedness and meaningfulness, and connectedness and pleasantness
* p < 0.05; ** p < 0.01; *** p < 0.001

|  | connectedness x meaningfulness | connectedness x pleasantness |
|---|---|---|
| music chosen by algorithm | 0.78<br>p < 2.2e-16 *** | 0.33<br>p = 3.316e-08 *** |
| music chosen by expert | 0.85<br>< 2.2e-16 *** | 0.53<br>p < 2.2e-16 *** |
| overall | 0.83<br>p < 2.2e-16 *** | 0.43<br>p < 2.2e-16 *** |

It could be argued, however, that we should be more interested in the correlation between connectedness and the change in meaningfulness and pleasantness after listening to music. A difficulty is that, since we are dealing with an ordinal scale, measuring change by the number of levels moved up or down a scale may not be legitimate. It is not legitimate, for example, to regard the transition from 'not at all' to 'not so' as being equivalent to the transition from 'very' to 'extremely', although they both involve a movement one step up the scale. However, if we put aside this reservation and calculate the change in meaningfulness and pleasantness, then we can calculate the correlation between these changes and connectedness, as shown in Table 16[21]. There is a positive correlation between connectedness and change in meaningfulness, and between connectedness and change in pleasantness. These correlations can be seen to be significant at the 95% level, since none of the 95% confidence intervals include zero. Given that, as noted in subsections 3.1.2 and 3.1.3, the effect of the music was frequently to diminish the meaningfulness and pleasantness, the tendency of the change in meaningfulness and pleasantness to correlate positively with connectedness perhaps suggests less that connectedness increases meaningfulness and pleasantness, but frequently mitigates their decrease. Finally, comparison of the two columns in Table 16 suggests that the effect of the connectedness is greater on meaningfulness than on pleasantness. This is particularly the case for the music chosen by algorithm, where there is no overlap between the 95% confidence intervals.

---

[21] Given that we are dealing with differences up and down a scale, and treating each inter-level distance as equal, we have calculated Pearson's correlation coefficient here.

Table 16. Pearson's correlation between reported connectedness and change in meaningfulness, and connectedness and change in pleasantness
* $p < 0.05$; ** $p < 0.01$; *** $p < 0.001$

|  | connectedness x change in meaningfulness | connectedness x change in pleasantness |
|---|---|---|
| music chosen by algorithm | 0.65<br>$p < 2.2e{-}16$ *** | 0.33<br>$p = 5.649e{-}08$ *** |
| music chosen by expert | 0.61<br>$p < 2.2e{-}16$ *** | 0.54<br>$p < 2.2e{-}16$ *** |
| overall | 0.64<br>$p < 2.2e{-}16$ *** | 0.44<br>$p < 2.2e{-}16$ *** |

### 3.1.5 Helpfulness

At the end of the study, participants were asked 'Overall, how helpful were the different pieces of music in developing an appreciation of the artwork?'. Figure 11 shows the distribution of responses, broken down by artwork and music groups. Kruskal-Wallis tests revealed no significant difference between the four artworks, neither for the music chosen by algorithm ($\chi^2(3, N = 66) = 3.9997$, $p = 0.2615$) nor that chosen by expert ($\chi^2(3, N = 72) = 2.7222$, $p = 0.4365$). A Mann-Whitney test showed that the music chosen by expert was significantly more helpful than that chosen by algorithm ($W = 1861$, p-value = 0.02418). Human expertise is likely to be important here; also the expert choices were drawn from a very wide knowledge of music, whereas the music chosen by algorithm was drawn from a much smaller database of 400 musical extracts. The greater reported helpfulness with the music chosen by expert may be related to the fact that the artworks were also judged more meaningful after listening to the music chosen by expert than after listening to the music chosen by algorithm; albeit in neither case was there a significant increase in meaningfulness after listening to music.

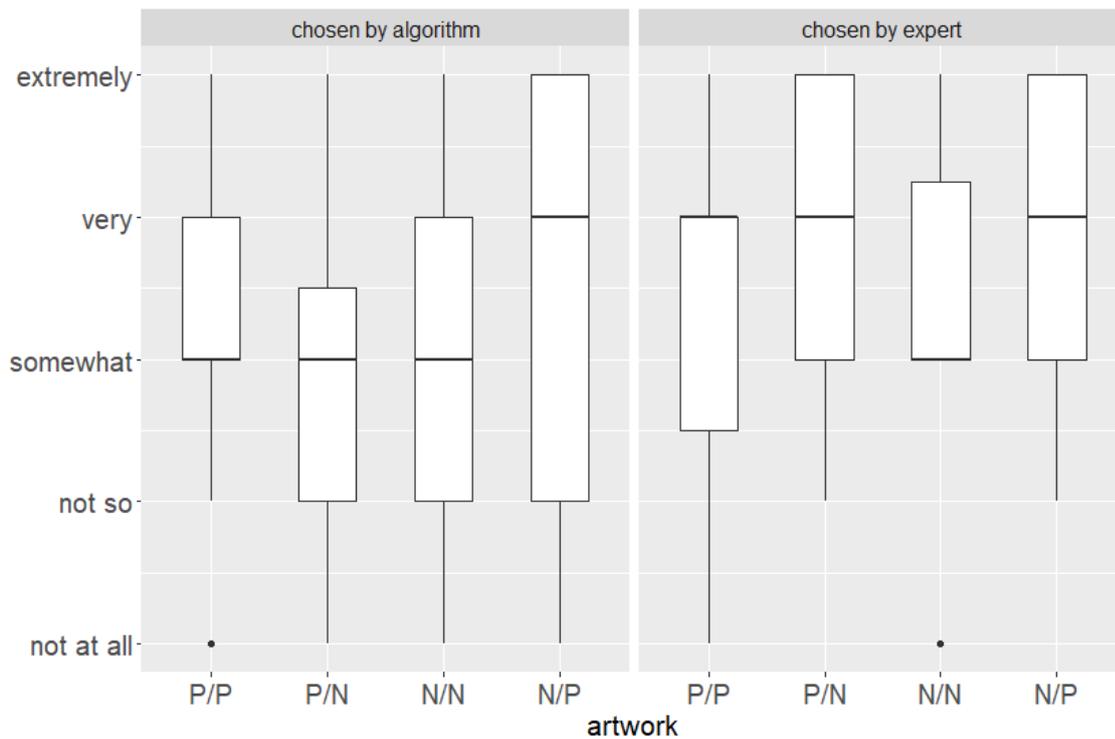

Figure 11. Helpfulness responses for each artwork and each music group.

The relatively high levels of helpfulness shown in Figure 11 may seem at variance with the generally negative effect which the music seems to have had on meaningfulness and pleasantness, as reported in subsections 3.1.2 and 3.1.3. One possibility is that this illustrates demand bias, i.e. the participants were responding how they thought the authors of the survey wanted them to respond. However, the significant difference between the two music groups does support the authenticity of the responses. It may be that participants rated helpfulness highly if only one or two extracts were helpful, i.e. they were not expecting all the musical extracts to be helpful.

Kruskal-Wallis tests indicated that helpfulness was not influenced by how often participants looked at art ($\chi^2(4, N = 138) = 1.792$, $p = 0.7739$), nor by how much time they spent listening to music ($\chi^2(4, N = 138) = 6.9956$, $p = 0.1361$). Nor did a Mann-Whitney test reveal any effect of gender (W = 1778.5, p-value = 0.5537).

However, there is an indication that helpfulness decreased with age, see Figure 12. Kruskal-Wallis tests showed that there was significant variation between the age-groups for the music chosen by algorithm ($\chi^2(5, N = 65) = 13.566$, $p = 0.01862$) but not for that chosen by expert ($\chi^2(5, N = 72) = 9.3972$, $p = 0.09423$). Jonckheere-Terpstra one-sided tests did reveal a decreasing trend, both for the music chosen by algorithm (JT = 661.5, p = 0.03687) and for that chosen by expert (JT = 726, p = 0.00377).

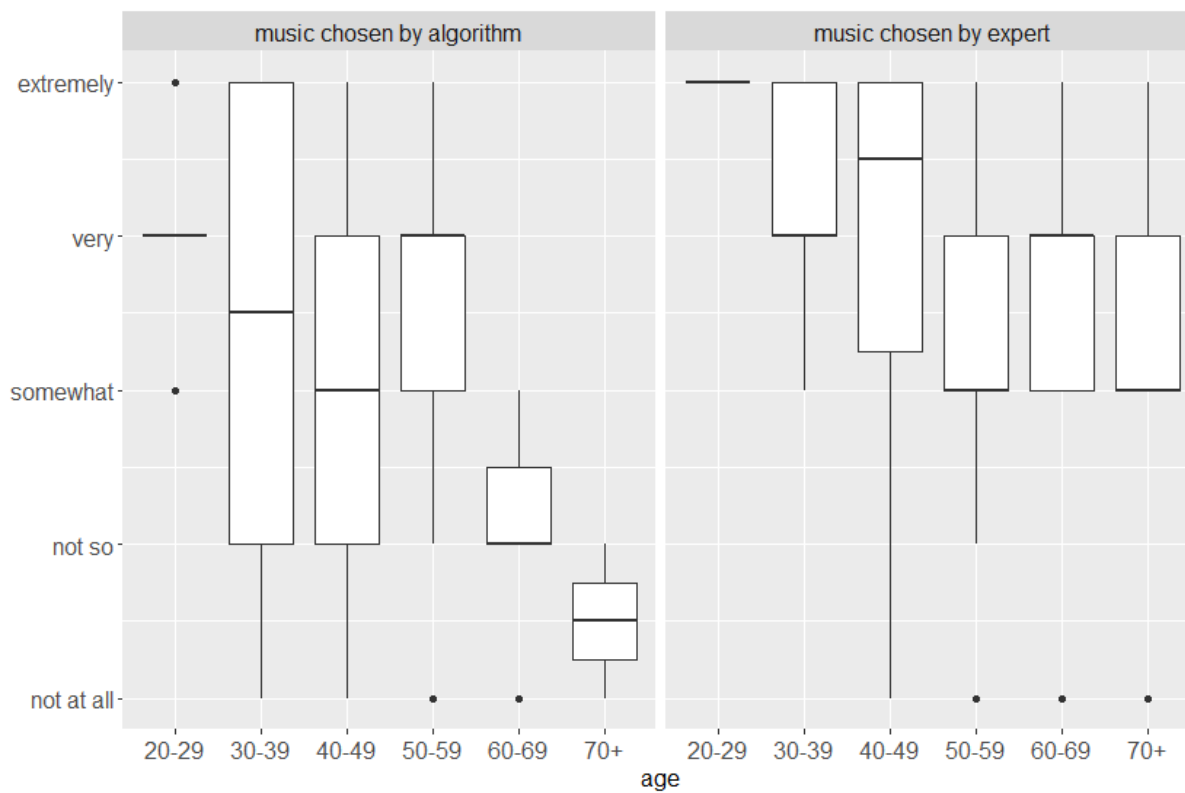

Figure 12. Helpfulness by age.

N.B. There were no participants under twenty and one participant listening to the music chosen by algorithm declined to say age. Also, as can be seen from Table 5 in subsection 2.5, there were relatively few participants at either end of the age spectrum, and this should be taken into account in interpreting the figure.

### 3.1.6 Summary

The evidence presented so far suggests that it is not possible to explain cross-modality in terms purely of induced or perceived emotions. Co-location of artwork and music in the same quadrant in the VA plane did not consistently lead to greater connectedness, nor to an increase in meaningfulness or pleasantness. Participants indicated a number of factors determining connectedness. For many artwork-music combinations emotion was the most frequently cited connection. However, this was not always the case, and other connections were frequently cited.

All significant effects of music on meaningfulness were negative, and this was also true for the majority of significant effects on pleasantness. Despite this, the majority of respondents did find the music helpful in developing an appreciation of the artwork.

Finally, the music chosen by expert was judged significantly more helpful than the music chosen by algorithm. This is consistent with the music chosen by expert being judged, overall, as more connected with the artworks; and with the artworks being judged more meaningful when heard with the music chosen by expert rather than with the music chosen by algorithm, although none of the musical extracts significantly increased meaningfulness.

## 3.2 Open-ended questions

In this section we look at the responses to the open-ended questions. We begin, in subsection 3.2.1 with the questions posed when participants first viewed the artworks, i.e. before listening to any music. In subsection 3.2.2 we discuss what participants noticed and felt after listening to the musical extracts. Subsection 3.2.3 continues the discussion of the previous subsection, but specifically looking the reactions to artwork and music collocated in the same quadrant of the VA plane. In subsection 3.2.4 we discuss what participants said about the connections they were making between artworks and music. In subsection 3.2.5 we discuss what participants had to say about the helpfulness of the music in appreciating the artwork, and what final comments they had. Finally, subsection 3.2.6 summarises our discussion.

### 3.2.1 Initial reactions to artworks

This section contained two open-ended questions. The first question was about what people noticed and thought about the artwork, the second about how they felt. Although, as we have already noted, some participants' responses were confounded, e.g. answering the second question as part of the first, here we separate out the two questions.

#### 3.2.1.1 Noticing and thinking

The first question was:
- What do you notice about the artwork? What are you thinking about the artwork?

Broadly, we can identify three aspects to participants' responses: content, style, and interpretation. Some participants concentrated on one of these aspects, whilst others discussed two or even all three.

As an example of a, relatively lengthy, response concentrating on content, one participant (3A5[22]) wrote, regarding *Festivities on the Coast* (P/P):

> "I first noticed the building and the way the sunlight was shining on it, and the shadows it created. I also noted how the sun was shining on the hills in the background although the sky is cloudy. It must be a hot day as there is a line of people dancing in the foreground and they are doing so in the shade. Also, some people are swimming or getting ready to swim. I wondered what the person was doing on the pole by the building."

As an example of a discussion of style, regarding *Hippopotamus and Crocodile Hunt* (N/P), a participant (4A10) wrote:

> "Exquisite skill and such detail in these paintings - the body muscles etc. Beautiful use of colours, the red linking up with the violent picture that was painted."

As an example of interpretation, combined with discussion of content, a participant 2N16 wrote about *Anachorète endormi* (N/N):

> "I feel conflicted - sorry for an old man who may be cold with few coverings but also peaceful watching a man sleep. Death is near (skull) and the pages of his life in the book is being reflected upon. There's a sunset also suggesting he's at the end of his life. The man is/was incredibly strong as he still retains his muscles and the

---

[22] In our coding of the unsupervised participants, the first character indicates the artwork viewed: 1 *Portrait of a Man*; 2 *Anachorète endormi*; 3 *Festivities on the Coast*; 4 *Hippopotamus and Crocodile Hunt*. The second character denotes whether the music was chosen by algorithm (A) or by expert (N). The subsequent number merely serves to identify the particular participant. Supervised participants are identified by an *S*, followed by a number.

*highlights of finger/toenails are fascinating. His forehead is untanned so when he worked it wasn't outside or he always wore a hat! The rough hessian of the cloth he's lying on suggests when we die our possessions are unnecessary."*

Finally, we provide an example of a participant (1A7) who, writing about *Portrait of a Man*, started with a comment on style, then referred to content, and concluded with some words about interpretation:

*"I notice the realism of the image. I feel a sense of calm contemplation. The image feels quiet and peaceful. The hands held in a prayer posture make me think that the subject is either in prayer or contemplating religion in some way."*

### *3.2.1.2 Feeling*

The second question was:
- How does the artwork make you feel?

Responses to this question were often relatively predictable. For example, writing about Festivities on the Coast, one participant (3A10) wrote:

*"It makes me feel happy, lively, wanting to go outside and join in."*

Another relatively predictable response was given by (4N10) in response to Hippopotamus and Crocodile Hunt:

*"It makes me sad. It gives me a sense of danger, urgency, excitement."*

Writing about *Anachorète endormi*, participant 2A11 wrote:

*"Sad, gloomy, uncomfortable."*

Another participant (2A9), writing about the same picture, started on a personal tone, but then went on to write more positively:

*"A little bit sad for the passing of time that comes to us all. He's in good physical shape though, despite his advancing years. So this makes me feel positively about the situation he is in."*

Some comments were less predictable, e.g. (2A7) also writing about *Anachorète endormi*:

*"Intrigued, curious. I kind of want to get to know him."*

Individual artworks were capable of eliciting a range of feelings, e.g. reacting to *Portrait of a Man*, participant 1N10 wrote:

*"Gloomy, distant, disconnected."*

Whilst participant 1N14 wrote, about the same artwork:

*"Makes me feel at ease, serene"*

### 3.2.2 Reactions after listening to music

The first two questions after listening to the music paralleled the two questions in the previous section, being concerned with what people noticed and felt differently. As before, the answers to these questions were confounded, but we separate them out here.

### *3.2.2.1 Looking differently*

Immediately after listening to each musical extract, the first question was:
- Did you now notice anything else about the artwork? Were you drawn to different things about the artwork?

In response to the first question, a few respondents found the music made no difference, e.g. participant 1A10 responded with 'no' after hearing each of the four pieces. Also, some participants found some of the musical extracts so distracting that they were not able to think about the artwork; e.g. participant 1N14 wrote, after listening to *Glass Dance Pieces No. 9*: *"no - the music is too fast paced to let my brain slow down to appreciate the artwork"*. However, most felt that, in some way, they were looking at the artwork differently. We divide these responses into three categories, although we are aware that the boundaries between them may not be distinct:

- seeing something in the artwork which the participant had not seen previously;
- focussing on a different part of the artwork, the implication being that this was something they were previously aware of, but that now they were giving it more focus;
- reinterpreting the artwork, or part of the artwork.

An objection can legitimately be made that perhaps these were the effects of looking again, and were not related to the music. Participant 4A2 was aware of this possibility:

> *"I am a bit sceptical though that being drawn to different things now as opposed to what I have already observed before is the result of the piece of music, since I now spent quite a bit of time viewing the picture and my mind is perhaps seeking new things to discover, with or without music."*

However, on many occasions participants' responses suggested their change in perception was being influenced by the music.

We now consider each of the three above categories in turn. Some people reported noticing something not noticed before. Participant 1A reported *"I noticed his crossed thumbs for the first time and was more drawn to the mountain outside the window"*, whilst viewing *Portrait of a Man* and listening to *Rubber*. Participant 2A9 reported noticing *"the body hair on his chest"*, whilst viewing *Anachorète endormi* and listening to *Cry*. Participant S6, viewing *Portrait of a Man* whilst listening to *Below*, noted being drawn *"very much to the expression on the man's face"*.

It is difficult to be confident, in cases like the those just reported, that it was really the music which led to something new being noticed. When we consider cases where participants' focus changed, then there is more evidence that this resulted from the music. An example of this is provided by participant 4N11 who, after viewing *The Hippopotamus and Crocodile Hunt* and listening to *Cry*, wrote *"The voices that sounded like horses neighing made me focus on the horses."* Participant 4A4, after viewing the same artwork, and listening to *Rubber*, wrote *"Focused more on the fear and the fight of the hippo and crocodile this time."* A more complex example is provided by participant 2A10 who, after viewing *Anachorète endormi* and listening to *Sunset*, wrote:

> *"Interestingly when the music started I noticed the black blanket/fur that covers the subject's lap. It seemed much blacker when the music was playing. I was drawn to the hands, something that I hadn't really paid much attention to before. I felt as if my vision was drawn into the centre of the picture and was sharper in that region. The outer regions of the picture became less focused and I wasn't as interested in looking at them."*

When we consider how participants reinterpreted the artwork, again there is evidence that this was genuinely the effect of the music. Participant 1A13, after viewing *Portrait of a Man* and listening to *Rubber*, wrote:

> *"I built more of a narrative about what was happening in the picture. It seemed like he was trying to make a decision and looking for guidance to help him on his resolution. The road and mountains in the background made me think that perhaps it was in relation to a journey or having to leave somewhere"*

Participant 4N4, after viewing *The Hippopotamus and Crocodile Hunt* and listening to *Spiegel im Spiegel*, commented:

> *"This really surprised me! I perceived movement in the painting! I noticed the clouds,*

*the grass and trees and the horses manes and I felt as though I was watching a scene in very slow motion."*

Participant 4N2 saw the same artwork in a less violent light when accompanied by *Dance Pieces No. 9*:

*"Interestingly with the music it felt a bit more like a cartoon, as if the people would get up and be fine and the animals would walk away unscathed."*

Participant S15 reported seeing *The Hippopotamus and Crocodile Hunt* almost in a different moral light, as a result of listening to *Spiegel im Spiegel*:

*"The piece of music definitely makes the piece of art overall feel more serene and less barbaric, in my opinion. For a brief moment, you start to see this as some noble battle that man has to fight to keep order in the world, rather than a savage attack on innocent animals. Perhaps taming and controlling the animals is the right thing to do?"*

On the other hand, music can make one view an artwork in a darker light. Participant 3N6 was influenced by *Threnody* to comment on *Festivities on the Coast*:

*"The artwork lost all of its 'original' meaning (assuming it had any) and it became a vehicle for a story of impending doom."*

### 3.2.2.2 Feeling differently

The second question was:
- How does the artwork make you feel with this piece of music? Is your feeling different from when you viewed the artwork without music?

Whilst the previous question was concerned with the artwork, and changes in its perception, our interest here is in the participants, and changes in their feelings.

We might expect that music with positive valence would have a positive effect on participants' feelings. This can take two forms. Where the initial reaction to the artwork was generally positive, we might expect an increase in positivity. An example of this is participant 1N4's reaction to *Portrait of a Man*. Initially, the participant reported feeling *"serene, quiet"*. After listening to *Dance Pieces No. 9*, the participant felt *"buoyant more positive"*. Additionally, where the initial feeling expressed negative valence, we might expect that positively valenced music would lead to a mitigation of that negativity. An example of this is offered by participant 4A3, who viewed *The Hippopotamus and Crocodile Hunt* (N/P). Initially, the reported feeling was *"Upset. I don't like hippopotomi or crocodiles but I don't like the idea that they're going to be killed like this."* After listening to *Sunset* (P/P), the participant reported *"For some reason I didn't feel as disturbed with the artwork when I listened to the music at the same time. It felt like I was watching characters in a video game."*

On the other hand, we would expect negatively valenced music to further depress a participants' feelings when viewing a negatively valenced artwork. Participant 4A8, after initially viewing *The Hippopotamus and Crocodile Hunt* (N/P), reported rather neutral feelings: *"Interested after a little while of looking. I didn't really feeling anything other than interested."* After viewing the artwork whilst listening to *Below* (N/N), the participant commented *"So at the start of the music I felt for the animals, I felt sad for them, empathy maybe."* Also, we would expect that negatively valenced music should at least reduce the positivity associated with a positively valenced artwork. An example of this occurred with participant 3N9, who viewed *Festivities on the Coast* (P/P). Initially, the participant reported feeling *"playful, but adrift from the action with the muted colours"*. After listening to Threnody (N/P), the participant's feelings were *"more intense, like something bad is happening."*

However, there were cases where the effect was not as might be expected. Participant 4A8, viewing *The Hippopotamus and Crocodile Hunt* (N/P), whilst listening to the positively-valenced *Sunset* (P/P), might have been expected to report an improvement in mood. Instead, the participant reported *"I really did not like this, it made me angry."* Perhaps there is too great a mismatch here between the emotion in the artwork and emotion in the music. Another example was provided by participant 3N13's reaction to *Festivities on the Coast* (P/P). Initially, the participant felt *"relaxed and in anticipation of something good, a nice surprise."* After listening to *Spiegel im Spiegel* (P/N), the participant wrote *"it added a bit of nostalgia, sadness, missing those times and these pastimes ... the feeling is of sadness and making you think of good times with people who are no longer with you."*

Sometimes the effect can be different from just a change in positivity or negativity of feeling. Returning to participant 1N4, who viewed *Portrait of a Man* (P/N), the feeling reported after listening to *Cry* (N/N) was *"Expectant as if something is about to happen"*. This response also illustrates the blurred distinction between what the participant feels and what the participant perceives; 'expectant' is a feeling in the participant, whilst 'something is about to happen' is a perception from the artwork.

Some participants reported a slowing down effect. This occurred for several participants with *Spiegel im Spiegel* (P/N). Participant 3N10, viewing *Festivities on the Coast* (P/P) wrote:

> *"There is now a timelessness about the picture. The building and hills have been there forever and always will be. The people may come and go but the society of which they are a part continues unchanged. They will meeting like this every year for eternity."*

Participant 4N9, viewing *The Hippopotamus and Crocodile Hunt* (N/P) and listening to the same music wrote:

> *"I could almost sense the animals moving around in a slow dance - almost a dreamlike state. I felt like my gaze was more drawn to the horses, perhaps as I would consider them a more graceful animal than the hippo or crocodile?"*

Participant S10, also viewing *The Hippopotamus and Crocodile Hunt* (N/P), but listening to *Dance Pieces No. 9* (P/P) had a similar experience:

> *"It make me feel more calm in the sense that I [am] even more [a] spectator ... the music make[s] put[s] you in a uh mental state in which you can calmly observe all the things without being disturbed by it ... battle scene is slowed down and you see very slow movements ... like someone who is there but who isn't there."*

In fact, the same participant, on listening to *Spiegel im Spiegel* also commented *"This music is slowing down time"*. It is probably significant that both the pieces of music mentioned as displaying this effect, *Spiegel im Spiegel* and *Dance Pieces No. 9*, have a repetitive nature.

### 3.2.3 Reactions to collocated artwork and music

We initially expected that there would be greater connectedness when artwork and music were in the same quadrant of the VA plane, and that this would lead to greater appreciation of the artwork. We have seen from subsection 3.1 that, based on responses to the multiple choice questions, this was not necessarily the case. We can complement the previous analysis by considering participants' reactions when artwork and music were collocated in the VA plane.

Examples of where collocation failed to increase the perceived pleasantness of the artwork can be seen by referring to Figure 10 and Table 13. As an example, consider the top left-hand plot, where the artwork and music has positive valence and arousal, and the music was chosen by algorithm. Here, the music collocated in the same quadrant has a negative effect on pleasantness; worse than two of the other pieces of music (P/N and N/N). Of this combination, one participant (3A6) wrote *"the music was too hectic"*. Another (3A7) wrote *"it felt out of place and jarring"*. The music sounded like disco-music, and participant 3A8 made some interesting comments:

> *"the 90s style ibiza music was so ill fitting it was laughable ... Completely mismatch and made me laugh ... Dance track with 1600s artwork is not compatible. Strangely if the instruments used were more in keeping with the artwork era the same melody may have created a huge connection."*

This illustrates a concern which occurred a number of times in the responses; participants were generally not happy when artwork and music were from different periods.

On the other hand, when we consider the top-right plot in Figure 10, where the music was chosen by expert, the collocated music did not significantly reduce the pleasantness, unlike for two of the other pieces of music. Consistent with this, some participants were positive in their reactions. Participant 3N8 wrote: *"I feel the music added a sense of joy, and expectation"*. On the other hand, participant 3N6 *"found the music distracting"*.

The artwork we have been considering so far in this subsection, with positive valence and arousal, was generally the best-received prior to listening to music; for that reason it might have been difficult to improve on the experience. If we consider the least well-received, which from Figure 10 appears to be the artwork with negative valence and positive arousal, then neither of the collocated pieces of music had a positive effect on pleasantness. With this artwork, after listening to the collocated music chosen by expert, participant 4N5 wrote *"dislike it more"*. Perhaps this is not surprising, collocation in this quadrant might mean that harsh music makes an unpleasant picture more unpleasant. Indeed participant 4A7 wrote *"This piece of music seems to intensify the imagery in the foreground of the artwork"*. Participant S15, who heard *Threnody*, wrote *"if anything it exaggerates my original thoughts"*. Participant 4A8 wrote:

> *"I feel that it started to put the picture into context. I looked at the animals more ... I felt as if to be in that moment, in the picture would be scary. I felt more alert when looking at the picture and I would argue that I enjoyed it more I felt like I was trying to understand it. Without the music I lacked interest in the image."*

Consistent with these comments, Figure 8 indicates that both the collocated music chosen by algorithm and that chosen by expert have the effect of increasing meaningfulness for this artwork, albeit not significantly.

### 3.2.4 Connections between artwork and music

Subsection 3.1.1 discussed the connections which people identified, out of a choice of four plus 'other', between the artwork and the music. This subsection is concerned with how they explained their answer.

#### *3.2.4.1 Emotion*

Overall, emotion was the most commonly indicated mode of connection, see the bottom right bar-chart in Figure 6, and a number of participants' comments referred to emotion. Participant 1N8, on viewing *Portrait of a Man* (P/N) whilst listening to *Spiegel im Spiegel* (P/N), commented:

> *"With the music there's a greater sympathy and empathy for the figure...almost starting to care more and connecting more with the emotional aspect of the painting."*

This was an example of artwork and music being collocated in the same quadrant of the VA plane. Another example of collocation, but in a different quadrant, was provided by participant 4A9, who viewed *The Hippopotamus and Crocodile Hunt* (N/P) and commented, after listening to *Rubber* (N/P):

> *"My response was purely emotional and as a result any cognitive rationality or coherent narrative did not happen in the timeframe."*

An emotional connection was also cited when artwork and music were in different quadrants. Participant 4N7, viewing *The Hippopotamus and Crocodile Hunt* (N/P) and listening to *Spiegel im Spiegel* (P/N) commented: *"The music brings up sad emotions."* This is interesting, because *Spiegel im Spiegel* has positive valence, albeit subdued by a negative arousal.

### 3.2.4.2 Theme and story

The next two commonly chosen modes of connection were theme and story. We discuss them in the same subsection because they sometimes appear interchangeable. Figure 6 shows that, overall, they were about equally popular, with slightly more participants opting for theme, caused by a greater preference for theme amongst those who heard the music chosen by expert. We shall, therefore, discuss theme first.

Participant 3N9, on viewing *Festivities on the Coast* (P/P) whilst hearing *Threnody* (N/P) cited emotion and theme as the connections, and wrote as explanation:

> *"A theme being there is a murder about to take place and no one knows surrounding [sic] as they're all going on their business."*

This is an example of the apparent interchangeability of theme and story. The quote could be interpreted as the participant developing a story, although story was not included as one of the connections. Theme and story were often cited together. For example, participant 1N10, who viewed *Portrait of a Man* (P/N) whilst listening to *Cry* (N/N), and cited theme and story, as well as emotion, wrote:

> *"The percussive, feel of the music made me think of impending battle or conflict. The young man possibly being aware of this then praying for rescue and safety."*

An example of a theme which does seem distinct from a story was provided by participant 2A6, whilst viewing *Anachorète endormi* (N/N) and listening to *Ping Heng* (N/P). The participant cited theme and emotion as connections, but not story, and wrote:

> *"the painting already has a religious atmosphere to it, but the music evokes religious vibes as well, reminding me of Gregorian chanting, and older religious music in general, while also evoking a sense of vast but enclosed space, and power."*

Sometimes, a theme can seem to relate to emotion. Participate 3N14, viewing *Festivities on the Coast* (P/P), whilst listening to *Dance Pieces No. 9* (P/P), cited both theme and emotion, and commented:

> *"Both the artwork and music were congruous with a celebration theme. They were both upbeat and joyous."*

Turning specifically to story as connection, there were cases where this was cited without citing theme. One example was provided by participant 1N2 when viewing *Portrait of a Man* (P/N) and listening to *Threnody* (N/P). The participant cited story and emotion as connections, and wrote:

> *"I felt like the man was facing his own death, perhaps through imminent execution and was resigned to it."*

Another example was provided by participant 3N9 when viewing *Festivities on the Coast* (P/P) and listening to *Cry* (N/N). The participant cited story and movement as connections and wrote: *"Made the people in the foreground feel more like a group, almost like a cult"*. The story here seems to be embryonic; one can imagine the idea of a cult being taken further and developed into a more developed story.

### 3.2.4.3 Movement

Overall, the fourth most commonly cited connection was movement. It is not surprising that, when we look at the rightmost column of Figure 6 we see that movement was more often a connection for the two artworks which themselves represented movement, i.e. *Festivities on the Coast* and *The Hippopotamus and Crocodile Hunt*. Participant 3N3 provided an example where only movement was cited as a connection. After viewing *Festivities on the Coast* (P/P) and listening to *Cry* (N/N), the participant wrote:

> *"I felt the music connected with the theme and title of the artwork. I could close my eyes and imagine a scene on the coast where people were indulged in revelry. A picnic on the coast with dancing, music and food. Where some were taking a dip in the warm waters whilst others were enjoying the day."*

Similarly, participant 4A9, after viewing The Hippopotamus and Crocodile Hunt (N/P) and listening to Sunset (P/P) cited movement as the only connection and wrote:

> *"I imagined the characters within the picture getting up and dancing, moving away from the scene and turning it into something happy."*

It appears that the participant made a connection between the movement, albeit of quite different sorts, in both artwork and music, whilst the positive valence of the music lifted the mood of the artwork.

It is interesting to see how participants found movement to be a connection when viewing two artworks where there was ostensibly no movement, and for which participants least frequently cited movement. Participant 1N10, after viewing *Portrait of a Man* (P/N) and listening to *Spiegel im Spiegel* (P/N) cited only movement as a connection, and then wrote: *"To me, both the music and the artwork conveyed a contrived sense of peace."* The participant appeared to be saying that the connection was an absence of movement. Similarly, participant 2N7, after viewing *Anachorète endormi* (N/N) and listening to *Spiegel im Spiegel* (P/N) cited only movement as connection, and wrote: *"Dreaming/slow passage of time/approaching the end of life."* Here, the connection is not an absence of movement, but rather a sense of moving slowly.

### *3.2.4.4 Other*

Finally, some participants responded with 'other'. Some combined other with one or more of the four suggested responses. However, there were 27 instances when participants responded only with other. Some of these were effectively saying they had found no connection, but some responses were more interesting. Participant S11, after viewing *Festivities on the Coast* (P/P) and listening to *Below* (N/N), found some aesthetic connection between artwork and music:

> *"I would say there is another type of connection here, it's more of an aesthetic connection, not so much an emotional one, because the emotional drive of the song might not fit the painting that much. It's more or less applied onto it while listening to the music, but there is this aura of not exactly festive and not exactly sad. Something in between that is happening in this picture. I would say there is a calmness to it. There is a there is a happiness and also a sorrow. I would say that is more of an aesthetic undertone to it, more than a specific thing happening in painting. I would say that's the best explanation I can give for that."*

Another participant (1A12), after viewing Portrait of a Man (P/N) and listening to Sunset (P/P), seemed to work hard to find a connection:

> *"Overall the music seemed at odds with the image, but maybe at a push there could be a link if you think of the combination as some kind of sci-fi story - the electronic music had elements of 1970s BBC style sci-fi, and maybe you could imagine the character in some kind of post-industrial agrarian future story, like The Tripods or something like that. But maybe i'm overthinking it."*

It should be added that the participant also clicked the box for no connection. Participant 2N8 also seemed to be working hard to find a connection; in this case between *Anachorète endormi* (N/N) and *Cry* (N/N). The participant wrote *"The [sic] was a detachment that forced you to look for reasons why the music was there."* The participant failed to see an emotional connection, even though both artwork and music were in the same quadrant.

Some participants used 'other' when one of the four suggested connections would have been appropriate. For example, participant 2A12, after viewing *Anachorète endormi* (N/N) and listening to *Below* (N/N) wrote *"the music made me think of an awakening."* Here, 'theme', i.e. of awakening, would have been a possible choice. Participant 2A1, viewing the same artwork and listening to the same music wrote *"it felt more soothing at first"*. Again, this might be classified as a shared theme, i.e. that of restfulness. Alternatively, particularly given that both artwork and music were in the same quadrant of the VA plane, this comment might be seen as a comment on a shared emotion. Similarly, participant 1N15, viewing *Portrait of a Man* (P/N) whilst listening to *Spiegel im Spiegel* (P/N) made a comment which implied a shared emotion *"The music was calming, this reflected my impressions of the picture."*

Rather similarly, participant 4N3, having viewed *The Hippopotamus and Crocodile Hunt* (N/P) and listened to *Cry* (N/N) wrote:

> *"There was a connection in the sense that both the scene and the music were unsettling, but chaotic in the case of the picture and discordant in the case of the music."*

This could be interpreted as an emotional connection created by the negative valence of both artwork and music.

Participant 3A4, after viewing *Festivities on the Coast (P/P)* and listening to *Sunset (P/P)* wrote *"I thought it was quite amusing to think what the dancers (?) in the foreground would have made of this piece of music."* This might have been classified under the shared theme of dancing.

Another example of what might be regarded as a shared theme was provided by participant 2N8 who, after viewing *Anachorète endormi* (N/N) and listening to *Dance Pieces No. 9* (P/P) wrote *"The music seems to express a finale - that this was an image of a person who had died having finally resolved the answer to his problems."* This can be seen to represent the shared theme of an ending; interestingly being constructed from artwork and music which are in opposing quadrants of the VA plane.

Participant 4A5 rather than finding a connection, seemed to find a lack of connection to be helpful. After viewing *The Hippopotamus and Crocodile Hunt* (N/P) and listening to *Rubber* (P/N), the participant wrote:

> *"I think the music was distracting enabling me to move away from the main image which was distasteful to me but probably a reflection of the era in which it was painted so probably helped to view the painting less emotionally"*

Here, the music seems to have deflected attention away from the unpleasantness of the image. Participant 4A3, viewing the same artwork whilst hearing Below (N/N) seemed to have the same experience:

> *"I felt I was viewing it somewhat dispassionately as if I was in a fugue state. Like I could see everything but I didn't have to feel the revulsion I initially felt because it 'wasn't about me'."*

An interesting response to the same artwork whilst hearing *Rubber* (P/N) was provided by participant 4A9:

> *"I felt a depth of involvement within the detail and the creation of the painting, the process itself, rather than the final image or story it depicts."*

The participant appears to have been diverted into thinking about the artwork as a constructed artefact, rather than a representation of a real-world image.

*3.2.4.5 Lack of connection*

Sometimes, lack of connection could be explained by distance between artwork and music in the VA plane. Participant 4N14, after viewing *The Hippopotamus and Crocodile Hunt* (N/P) and hearing *Dance Pieces No. 9* (P/P) wrote *"The music was fast paced and added a sense of urgency and excitement but this didn't connect to the image which was more brutal, distressful and sad."*

At other times there was a lack of connection, when positioning in the VA plane might predict an emotional connection. Participant 1A1, after viewing *Portrait of a Man* (P/N) and listening to *Rubber* (P/N), wrote *"I felt there was a juxtaposition between the music and artwork which made me uncomfortable."*

Participant 2N15 provided another example where artwork and music were in the same quadrant, but there was a failure to connect. After viewing *Anachorète endormi* (N/N) and listening to *Cry* (N/N), the participant wrote

*"I cannot identify any connection between the listened musical piece and the visual composition. In my opinion, the music did not resonate at all with the image conveyed by the artist. The person in the painting appears to be asleep, presumably after a long period of work, and the artist portrays a sense of stillness. Meanwhile, the music seemed to evoke action."*

This is a particularly interesting comment, since both artwork and music had negative valence and, in particular, negative arousal. Despite this, the participant clearly felt there was a difference in energy between the artwork, representing "stillness", and the music, representing "action". This raises the question as to how well *Cry* represents the bottom left corner of the VA plane. We return to the general question of allocation of artwork and music to quadrants when we discuss limitations of the study in subsection 4.1.

Some participants explained why they did not indicate any form of connection. A number of participants were aware that the artwork and music were anachronistic. Participant 3A9, viewing *Festivities on the Coast*, noted anachronisms both after hearing *Rubber* (*"the music ... was too 'modern' to fit"*) and *Sunset* (*"the music was just too electronic and modern to fit with the theme of the artwork"*). Participant 3A13, after viewing the same artwork and listening to *Rubber*, made a similar comment: *"The music was modern and the painting isn't."* Participant 1A11, after viewing *Portrait of a Man*, and listening to *Rubber*, wrote *"I felt that I struggled to connect the music with the artwork, it felt like the two pieces are from very different time periods"*.

### 3.2.5 Helpfulness and final comments

In the final part of the study participants were asked two open-ended questions. Firstly, having rated the helpfulness of the music in developing an appreciation of the artwork, they were asked to explain their answer. Secondly, they were asked if they had any other comments about the experience.

*3.2.5.1 Helpfulness*

Consistent with the results shown in Figure 11, participants generally found at least some of the music helpful, although as participant 2N17 observed, it does depend on the combination of artwork and music:

> *"It really depends on the music and the image. Do they complement each other? Otherwise, it can distract rather than enhance the experience of looking at the artwork."*

An interesting view of how different pieces of music have different effects was given by participant 2A7, who viewed *Anachorète endormi* (N/N) and listened to the music chosen by algorithm:

> *"Three of the pieces gave me a different mindset, one didn't fit at all. The three that I thought fit changed my thinking of what I was seeing in terms of the thematic interpretation between death, very strong in the choral type of music, and sleep, dominant in the more rhythmic pieces."*

Participant 4A5, who viewed *The Hippopotamus and Crocodile Hunt* (N/P) in conjunction with the music chosen by algorithm, implied that the music might have an indirect effect, in that it could encourage viewing for longer:

> *"I think the music enabled me to remain engaged with a painting which I would normally steered away from quickly however my focus remained on the story of the painting and my dislike of the style remained unchanged."*

Whilst most comments were generally favourable, there were some dissenting voices. Participant 3A6, after viewing *Festivities on the Coast* (P/P) and listening to the music chosen by expert wrote:

> *"I thought the artwork on its own could be appreciated for what it was, in terms of technique, choice of themes etc. The music, any music, made it feel like part of something else and assigned meaning post hoc (and ad hoc!). The only music that could make me more immersed in the painting itself would be secular street music from the 17th century. Anything else gets me thinking about all sorts of scenarios that have nothing to do with the painting itself, and essentially downgrade the image to a prop for a story."*

This raises the concern that the music is imposing an interpretation on the artwork which would not be there otherwise, and may well not be an interpretation intended by the artist.

Many participants thought that, for good or ill, music had a strong effect on the experience of the artwork. Participant 3A7, who viewed *Festivities on the Coast* (P/P) and listened to the music chosen by algorithm had a mixed reaction to the various pieces of music:

> *"I hadn't realised how influenced I might be by music as I viewed a piece of art. It was a revelation to me."*

### *3.2.5.2 Final comments*

Some final comments reflected a very positive experience. Participant 4A6 wrote *"Some of the music was really well chosen, but they all brought out different elements of the painting."* Participant 2A8 was even more favourable *"I think I may actually make an effort to go to an art installation if appropriate music was played. It has made me really look at the painting and try to appreciate it in more depth."*

Participant 4A14 was similarly positive about the use of music in galleries:

> *"It's such an effective experience I think I would appreciate art galleries a lot more if they had this added dimension. It makes art more visceral, more three dimensional and experiential."*

Some participants were surprised by how much the music could affect the experience of looking at art, e.g. participant 3A9 wrote:

> *"It was a really interesting experiment, I enjoy listening to music and listening to a wide variety depending on my mood. I didn't realise how much it can affect how we see things. Thank you for letting me take part."*

However, other comments were less favourable. Participant 3N6 gave a mixed response *"It was more annoying than I thought it would be, because I kept expecting a sort of historically appropriate track that never came! But it was very interesting."* This is consistent with comments reported above where participants found the artwork and music anachronistic; although in this case the participant had not identified any particularly anachronistic musical extract.

Participant 3N10, having viewed *Festivities on the Coast* (P/P) and listened to the music chosen by expert echoed the comment made by participant 3A6 and reported in the previous subsection:

*"I feel that using music like this is dangerous because it risks subconsciously imposing an interpretation different from that intended by the artist."*

Participant 3A10 had similar views:

*"My other observation is that I felt silence was the best way to appreciate this, each music piece seem to impose itself and make it harder for the art work to tell its own story and let details come to live in their own time. It was like being dictated what to feel, rather than being allowed to appreciate the work through its own energy and story."*

Participant 4A9 made a balanced comment, noting the advantages and disadvantages:

*"It is interesting to note there is a fine line between the music enhancing the experience of the artwork, dominating it or detracting from it."*

Participant 4N10 stressed the multi-modal aspect of the experience, writing:

*"It is interesting to bring another sense into art appreciation. I know that enjoyment of art is subjective, and to add another layer of subjectivity in terms of musical taste is thought-provoking."*

Even a participant, 1N7, who was not particularly supportive of combining artwork and music, wrote *"It was interesting, overall it confirmed why I like to view art in quiet spaces :)"*.

### 3.2.6 Summary

Participants' reactions to the study experience were wide-ranging. Many participants viewed the artworks differently as they listened to the different pieces of music, either noticing new features, focussing on different aspects of the artworks, or interpreting the artworks differently. Whilst reinterpretation of the artwork appeared frequently to be directed by the music, it may be argued that noticing or focussing on different features could, at least in some cases, be caused simply by additional time viewing the artwork. Even if this is the case, the presence of music may encourage longer viewing.

Participants feelings were also altered by the music. Often, the effect was as one would predict from the valence of the music, although there were counterexamples to this. In particular, collocation of artwork and music in the same quadrant of the VA plane did not always enhance the experience. This may have been because of a mismatch in style, e.g. the music being too hectic for the artwork, or a sense of anachronism. An interesting effect reported by several participants was that of a slowing down of time. This occurred with two of the pieces, *Dance Pieces No. 9* and *Spiegel im Spiegel*, perhaps because of the repetitive nature of the pieces.

When asked to explain the connections made between artwork and music, the explanations were largely predictable. Some comments did evidence a blurring of the boundaries between the four predefined options. One participant cited *story* when *theme* might have been more appropriate; alternatively, the story could be regarded as embryonic. Sometimes two modes of connection were reflecting the same experience. Another participant, cited both *theme* and *emotion*; for the former specifically referring to a celebration theme and talking also about the artwork and music being 'upbeat and joyous'.

Where participants had indicated 'other', their explanations sometimes suggested they could equally have used one of the four predefined options. Sometimes participants seemed to work hard to find a connection, perhaps feeling obliged to look for one.

Participants often gave an explanation for a lack of connection. Although such participants may not have referred to emotion, at times their explanations could be interpreted as an emotional mismatch, i.e. caused by distance in the VA plane. At other times, proximity in the VA plane would have predicted a connection, but the participant found none. Sometimes this was because of an anachronistic relationship between artwork and music. At other times, the reason was less clear-cut. It may be, that in some cases allocation to VA quadrant was not always accurate.

As discussed in subsection 3.1.5, many participants found the experience helpful in appreciating artwork; although this seems surprising in view of the analysis of change of meaningfulness and pleasantness reported in subsections 3.1.2 and 3.1.3. The comments in the final section of our study reported a range of reactions. There were some participants who wanted to appreciate art in silence. There were others with a more mixed approach, appreciating the contribution of some of the music extracts whilst reacting against the other extracts.

Each of the 138 participants gave oral or written responses to 16 open-ended questions, making a total of 2208 pieces of text. We have attempted to provide a representative sample of those responses, although not every viewpoint may be fully represented. In the next subsection we take a more systematic approach, applying sentiment analysis to participants' responses.

## 3.3 Sentiment analysis

In this subsection we use sentiment analysis to investigate the responses to some of the open-ended questions we posed to participants. Our object is to determine to what extent sentiment analysis confirms, or adds to, the analysis of the preceding subsections.

In subsection 3.3.1 we discuss our approach to sentiment analysis. In subsection 3.3.2 we discuss how the valence and arousal of the initial text extracts, i.e. before listening to music, depended on the valence and arousal of the artworks. In subsection 3.3.3 we extend this analysis to the responses after listening to the music, including investigating the change in valence and arousal of the text. In subsection 3.3.4 we analyse participants' explanations of their ratings of the helpfulness of the music in developing an appreciation of the artwork. Finally, in subsection 3.3.5 we summarise our findings.

### 3.3.1 Method

Subsections 2.1 and 2.2 explained how we used sentiment analysis to select the artworks and one of the groups of music. In particular, we explained how we used the lexicon prepared by Mohammad (2018) to assign valence and arousal values to artwork and music, based on emotion-describing words associated with a range of possible artworks and musical extracts. We take the same approach here to assign valence and arousal to participants' responses. However, previously we were analysing sets of tags, generally adjectives or nouns, whilst we now have to deal with words in sentences. For example, we need to identify a phrase such as 'not good' as having a quite different emotional content from 'good'; see the discussion in Kiritchenko and Mohammad (2017). We have adopted a relatively simplistic approach. We assign valence and arousal to words in the text, using the Mohammad (2018) lexicon[23], However, where a word is preceded by a negator, such as *not*, we negate the value of the valence, although leaving the arousal unchanged. The full set of negators used were[24]:

*no, not, never, dont, don't, cannot, can't, won't, wouldn't, shouldn't, aren't, isn't, wasn't, weren't, haven't, hasn't, hadn't, doesn't, didn't, mightn't, mustn't*

Additionally, before taking account of the negators we remove the words *a, the, an, some, much, such*. This is to allow for possible phrases such as 'not a good'.

Finally, we calculated the mean valence and mean arousal of each textual response, ignoring words not in the lexicon, i.e. in calculating the means, we use the number of lexicon words in the text, including repetitions, not the total number of words.

### 3.3.2 Response to initial questions

As a reminder, the first two questions, after initially viewing the artwork, were:
- What do you notice about the artwork? What are you thinking about the artwork?
- How does the artwork make you feel?

We merged the responses to the two questions and estimated the valence and arousal of each response using the approach described in subsection 3.3.1. Table 17 shows the mean valence of the responses for each of the four artworks.

Table 17. Mean valence of responses to first two questions, prior to hearing music; 95% confidence interval in brackets

|  | artwork valence negative | artwork valence positive |
|---|---|---|
| artwork arousal positive | 0.02 (-0.05, 0.07) | 0.33 (0.29, 0.38) |
| artwork arousal negative | 0.14 (0.07, 0.20) | 0.32 (0.28, 0.36) |

The first point to note is that all these means are positive. In fact, of the 138 pieces of text, only 17 had non-positive values. A type III Anova[25] revealed significant effects of artwork valence ($F(1, 134) = 83.8234$, $p = 7.998e-16$) and of artwork arousal ($F(1, 134) = 4.4760$, $p = 0.03622$), and also a significant interaction effect ($F(1, 134) = 6.2003$, $p = 0.01400$). One summarisation of Table 17 is to note that the text valence is positively correlated with the artwork valence; and that when artwork valence is positive, then artwork arousal has no effect, whereas when artwork valence is negative, a positive artwork arousal has the effect of further decreasing text valence.

Table 18 shows the mean arousal of the responses for each of the four artworks. A type II Anova revealed significant effects of artwork valence ($F(1,134) = 55.7835$, $p = 9.307e-12$) and of arousal ($F(1,134) = 84.0804$, $p = 7.384e-16$), but no interaction effect ($F(1,134) = 0.1267$, $p = 0.7224$). Because of the lack of interaction, we can summarise Table 18 by noting that text arousal correlates more-or-less independently with artwork valence and artwork arousal. For artwork arousal the correlation is positive; for artwork valence, the correlation is negative. In summary, artwork arousal stimulates text arousal, whilst artwork valence dampens text arousal. This subject to a similar caveat to that made in subsection 3.1.1. We have only one artwork from each quadrant of the VA plane and cannot be certain to what extent factors other than valence and arousal are having an effect. We return to this subject in Section 4.

---

[23] We extended this lexicon slightly. Specifically, we included *better* and *best*, setting their valence and arousal values to that of *good*; and also *worst*, setting its values equal to *worse*.
[24] From Zhang (2018), see https://books.psychstat.org/textmining/sentiment-analysis.html#gram-sentiment-analysis
[25] Type 3 was used because there was a significant interaction effect.

Table 18. Mean arousal of responses to first two questions, prior to hearing music; 95% confidence interval in brackets

|  | artwork valence negative | artwork valence positive |
|---|---|---|
| artwork arousal positive | 0.01 (-0.03, 0.05) | -0.15 (-0.17, -0.11) |
| artwork arousal negative | -0.18 (-0.21, -0.11) | -0.32 (-0.36, -0.29) |

### 3.3.3 Responses after listening to music

Immediately after listening to each musical extract, participants were asked two questions analogous to those discussed in the previous subsection:
- Did you now notice anything else about the artwork? Were you drawn to different things about the artwork?
- How does the artwork make you feel with this piece of music? Is your feeling different from when you viewed the artwork without music?

We merged the responses to these two questions and performed sentiment analysis in the same manner as before. Figure 13 shows the text valence before listening to music, and after listening to each extract. Figure 14 similarly shows the text arousal. Unlike in the previous subsection, text valence and arousal prior to listening to music are presented separately for each group of participants, to enable comparison with valence and arousal after listening to the music. As one would expect, the valence and arousal of the text prior to listening to music are similar for both participant groups, and follow the pattern discussed in the previous subsection. Both figures also show the 95% confidence intervals. These should be taken into account in interpreting the plots.

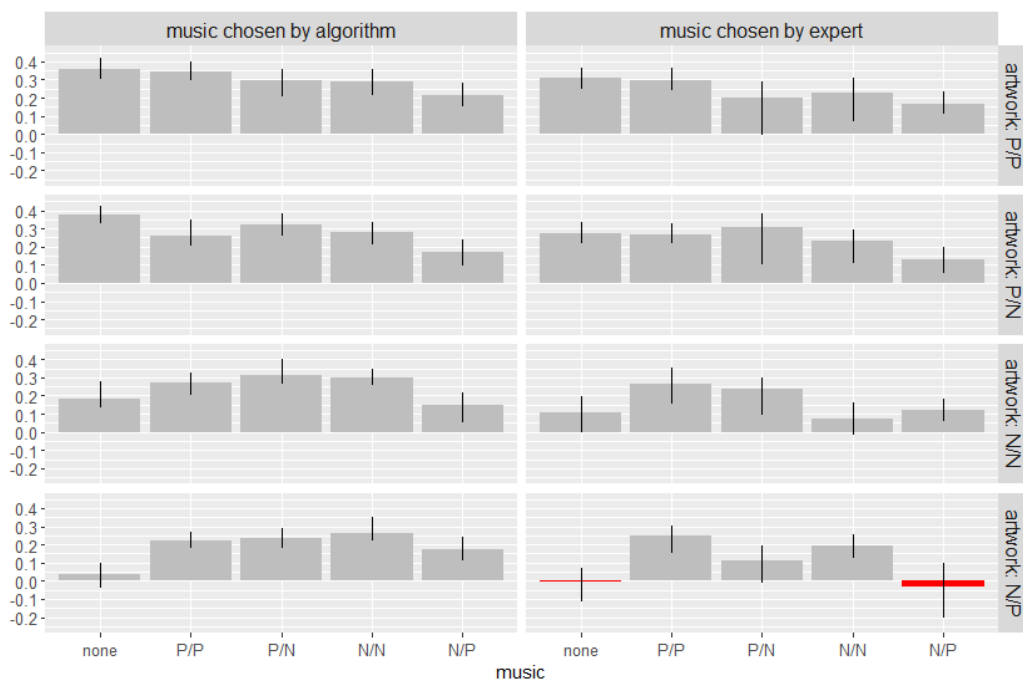

Figure 13. Text valence for first two questions without music, and for each artwork-music combination; negative values in red; lines show 95% confidence intervals

For the text valence and arousal after listening to each piece of music, there are five independent factors to consider:
- the valence and arousal of the artwork;
- the valence and arousal of the music;
- the particular group of musical extracts used.

We consider first the text valence. A five factor ANOVA shows a significant dependence of text valence on music group (F(1,520) = 14.5936, p = 0.0001495); the mean text valence when the music was chosen by algorithm was 0.26, with 95% confidence interval (0.24, 0.27) whilst the mean text valence when the music was chosen by expert was 0.19, with 95% confidence interval (0.17, 0.22). Moreover, there were also a number of significant interactions involving the music group. These suggest that there are factors which differ between the two music groups and influence the valence of the text.

The significant effect of music group, and its significant interactions with other factors make interpretation of the five factor ANOVA hard to interpret. Therefore, we performed a four factor ANOVA on each of the two music groups separately. For the group with the music chosen by algorithm, the only significant factor was the arousal of the music (F(1,248 = 9.4769), p = 0.002315), and there were no significant interactions. Music arousal had a negative effect on text valence; the mean text valence in the cases of positive music arousal was 0.23, with 95% confidence interval (0.20, 0.25), whilst the mean text valence in the cases of negative music arousal was 0.29, with 95% confidence interval (0.27, 0.31). This effect is discernible in Figure 13. For each artwork, the music with negative valence and positive arousal gave rise to the lowest text valence; whilst the music with positive valence and positive arousal gave rise to the second lowest text valence, with the exception of the artwork with positive valence and arousal, i.e. the top row.

A four factor ANOVA on the group with the music chosen by expert gave a much more complex picture. There were two significant factors: artwork valence (F(1,272) = 6.0768, p = 0.014315) and music valence (F(1,272) = 6.5031, p = 0.011317). In both cases, Figure 13 indicates the effect is positive. However, there were four interaction effects, between:
- artwork arousal and music valence (F(1,272) = 6.2637, p = 0.012911);
- artwork arousal and music arousal (F(1,272) = 7.8564, p = 0.005429);
- artwork arousal, artwork valence, and music arousal (F(1,272) = 5.1685, p = 0.023779);
- artwork arousal, music valence, and music arousal (F(1,272) = 7.6668, p = 0.006011).

It is difficult to interpret these interactions. However, we can see from the right-hand side of Figure 13 that there are perturbations from the general trend of increasing text valence with increasing artwork and music valence. For example, considering the bottom plot (artwork N/P) in Figure 13, we see that the text valence after listening to music P/N is considerably less than after listening to music P/P[26].

Figure 13 shows very clearly that for the artworks with negative valence, and particularly for the artwork with negative valence and positive arousal, in general the effect of the music is to considerably raise the text valence; in these two cases, participants' responses after listening to the music are considerably more positive than on first viewing.

We might expect that text valence bears some correlation to the reported pleasantness of the artwork, as shown in Figure 10. Indeed, there are some similarities in shape between Figures 10 and 13. Considering all five experiences of viewing the artwork for each of the 138 participants, i.e. a total of 690 data items, the Spearman rank correlation between text valence and reported pleasantness was 0.36, with 95% confidence interval (0.29, 0.42)[27]. This correlation is 0.66 (0.56, 0.74) when calculated on the responses prior to listening to music; 0.22 (0.10, 0.34) calculated on the responses after listening to the music chosen by algorithm; and 0.36 (0.25, 0.46) calculated on the responses after listening to the music chosen by expert.

---

[26] We should also point out that the sample sizes represented by each bar in Figure 11 are small, i.e. between 15 and 20.

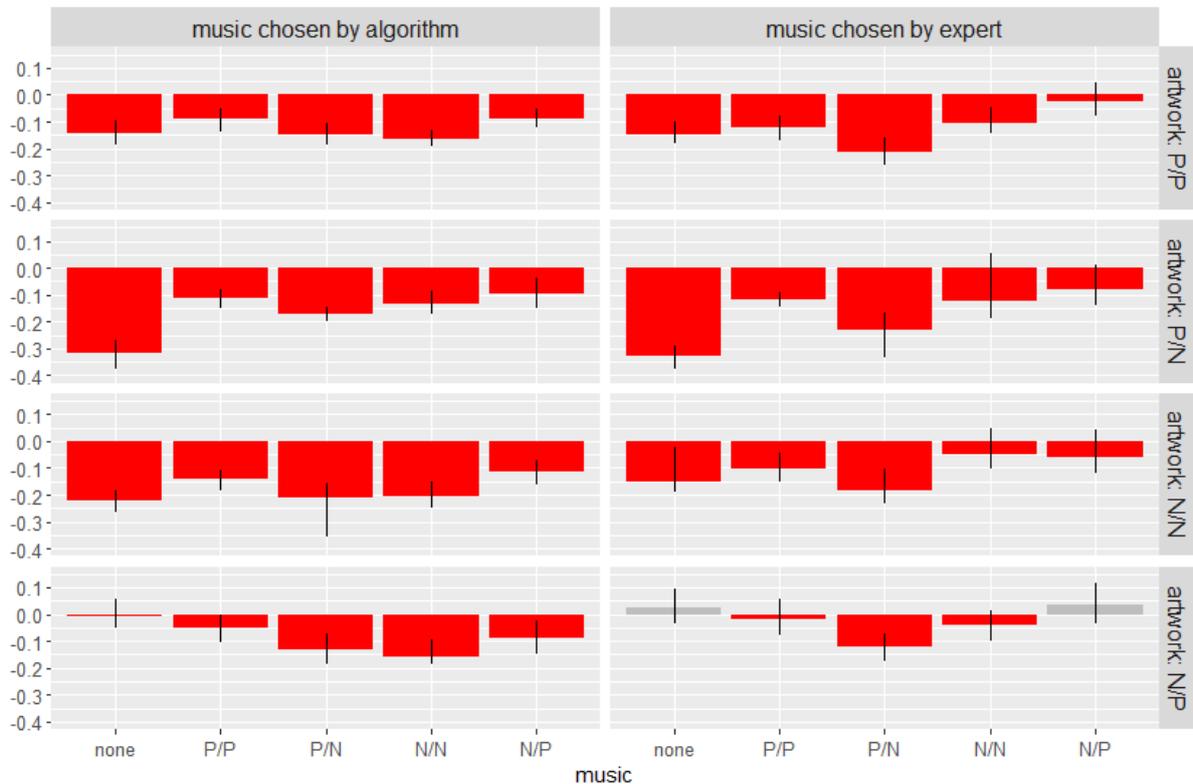

Figure 14. Text arousal for first two questions without music, and for each artwork-music combination; negative values in red; lines show 95% confidence intervals

Considering text arousal, the first point to note is that, with two exceptions, all the mean text arousal values shown in Figure 14 are negative. The two exceptions are very small positive values, occurring in the bottom right plot. Secondly, as we would expect, the values prior to hearing the music are similar for both groups of participants, and follow the pattern discussed in subsection 3.3.2. As a result, for both groups, the most negative text arousal occurs with the artwork with positive valence and negative arousal, and the least negative text arousal occurs with the artwork with negative valence and positive arousal. For these two artworks, each of the music extracts has the effect of reducing the negativity of the most negative and increasing the negativity of the least negative. For the other two artworks, the effect of the music is more varied.

A five factor ANOVA reveals a significant dependence on the music group ($F(1,520) = 13.2071$, $p = 0.0003066$). The mean text arousal when the was music was chosen by algorithm was -0.13, with 95% confidence interval (-0.14, -0.12); whilst the mean text arousal when the music was chosen by expert was -0.10, with 95% confidence interval (-0.12, -0.08). There were also two significant interaction effects involving the music group; therefore, we shall follow the same procedure as with text valence, and treat the two music groups separately.

---

[27] We use Spearman's rank correlation because pleasantness is an ordinal variable, although text valence is a continuous variable. Also, if we consider not the original data points, but the median pleasantness and mean valences as displayed in Figures 10 and 13, then the Spearman's rank correlation is 0.67 with 95% confidence interval (0.46, 0.81).

For the music group chosen by algorithm, there are two significant factors, artwork valence (F(1,248) = 4.2870, p = 0.03944) and music arousal (F(1,248) = 5.8398, p = 0.01639), and no significant interactions. The effect of music arousal is easy to discern from Figure 14; for each artwork, i.e. row, the two musical extracts with positive arousal (the second and fifth in each row) give rise to less negative text arousal than the other two pieces (the third and fourth in each row). The effect of artwork valence is less pronounced. For positive artwork valence, the mean text arousal is -0.126, with confidence interval (-0.141, -0.110); for negative artwork valence, the mean text arousal is -0.137, with confidence interval (-0.160, -0.117).

For the music chosen by expert, the only significant factor was the music valence (F(1,272) = 8.5188, p = 0.003809). From Figure 14 we can see that positive valence tends to increase the negativity of the text arousal.

### 3.3.4 Helpfulness

In subsection 3.1.5 we discussed how participants rated the helpfulness of the music in developing an appreciation of the artwork. In this subsection, we present a sentiment analysis of the textual explanation of their answer. Table 19 shows, for each of the four artworks, the median helpfulness, mean text valence and mean text arousal of this textual answer. Both the median helpfulness and the mean text valence suggest that the music may have been more helpful with the two artworks with positive arousal. However, Kruskal-Wallis tests in subsection 3.1.5 indicated no significant difference in helpfulness between the artworks. On the other hand, a two factor Type II ANOVA did reveal a significant dependence of text valence on artwork (F(3,130) = 3.8647, p = 0.01095), but no significant dependence on the music group (F(1,130) = 0.0892, p = 0.76572) and no significant interaction effect (F(1,130) = 0.5070, p = 0.67810). When we compare the artworks with positive arousal and those with negative arousal, a Mann-Whitney test indicates no significant dependence of helpfulness on artwork arousal (W = 2612.5, p = 0.2729). However, a t-test indicated a significant dependence of the text valence on artwork arousal (t(126.79) = 3.015, p = 0.003105), as is suggested by Table 19. It would appear that, for both music groups, participants responded more positively to the question about helpfulness of the music when viewing the artworks with greater arousal, i.e. *Festivities on the Coast* and *The Hippopotamus and Crocodile Hunt*.

We also investigated the dependence of the text arousal on artwork and music group. A two-factor ANOVA indicated no significant dependence on either of these factors, and no significant interaction.

Table 19. Helpfulness responses versus artwork valence and arousal; figures in brackets indicate 95% confidence intervals

|  | negative artwork valence | | | positive artwork valence | | |
|---|---|---|---|---|---|---|
|  | median helpfulness | mean text valence | mean text arousal | median helpfulness | mean text valence | mean text arousal |
| positive artwork arousal | 3 | 0.38 (0.34, 0.41) | -0.13 (-0.17, -0.10) | 3 | 0.37 (0.34, 0.41) | -0.10 (-0.13, -0.06) |
| negative artwork arousal | 2 | 0.30 (0.27, 0.33) | -0.14 (-0.16, -0.11) | 2 | 0.34 (0.31, 0.37) | -0.12 (-0.15, -0.09) |

### 3.3.5 Summary

On initial viewing of the artwork, the valence of the textual response was significantly positively affected by the artwork valence and negatively affected by the artwork arousal, with a significant interaction. As a result, artwork arousal has little effect when the valence is positive, but a considerable negative effect when the artwork valence is negative, i.e. positive arousal amplifies the effect of negative valence. The arousal of the textual response was significantly positively affected by the artwork arousal and significantly negatively affected by the artwork valence, with no significant interaction effect; i.e. artwork arousal stimulated arousal in the text, whilst artwork valence depressed the textual arousal.

After listening to the music, both text valence and arousal were significantly different for the two music groups, with significant interaction between music group and the other factors. For the music chosen by algorithm, music arousal had a significant negative effect on text valence, and there were no other effects. Music arousal had a significant positive effect on text arousal, and artwork valence had a significant negative effect. For the music chosen by algorithm, the situation is more complex for text valence; there is a general trend of increasing text valence with increasing artwork and music valence, with perturbations corresponding to the interactions. For text arousal, the only significant factor was music valence, which depressed the text arousal. The different behaviours for the two music groups seems to undermine any attempt to explain text sentiment of the helpfulness responses in terms of artwork and music sentiment. However, for the music chosen by expert, the assignment to quadrants in the VA plane were performed by one person, and may not have corresponded completely to the perception of participants. We return to this point in subsection 4.1.

We also note that there was a significant positive correlation between the text valence and the reported pleasantness of the artwork. This was appreciably higher when calculated from the responses prior to listening to the music than when calculated from the responses after listening to music.

Finally, whilst the response to the question about the helpfulness of the music revealed no significant differences between the artworks, an analysis of the explanation given by participants revealed that the text valence was significantly higher for the two artworks with positive arousal than for the other two. This suggests that the music was more helpful in developing an appreciation of the artworks with high arousal.

## 4 Discussion

In subsection 4.1 we discuss some limitations of our study and some considerations for future experimental design to overcome those limitations. In subsection 1.4 we posed four research questions. Subsection 4.2 answers the first three of those questions by summarising our key findings. Subsection 4.3 answers the fourth of those research questions by presenting some guidelines for using cross-modality to enhance the aesthetic experience of art. Finally, subsection 4.4 proposes some future areas for research.

### 4.1 Limitations of the study and considerations for future experimental design

This study has been an exploratory one. We believe it helps to map out the terrain for future research, but has a number of limitations which prevent a definitive interpretation of certain of the results. For our study, we only used one artwork and two musical extracts to represent each quadrant of the VA plane. This fails to compensate for confounding factors, i.e. factors other than the sentiment in the music and artwork. Ideally, a future study could use more artworks and musical extracts to average out the effects of confounding factors. At the same time, more care could be taken to avoid confounding factors. Our four artworks were quite different. Two were portraits of single individuals and two were groups of individuals. Even for the latter there was a difference; *The Hippopotamus and Crocodile Hunt* was a close-up of men and animals, whilst *Festivities on the Coast* was a land- and seascape including more people, but at a distance. Ideally we should avoid, e.g. differences of genre. On the other hand, it may be argued that some of the apparently confounding differences are really necessary to obtain the required sentiment, e.g. it may be difficult to obtain high arousal with a portrait of a single person. This objection might be able to be overcome. One could imagine four pictures, each with two people. For positive valence and arousal they might be dancing vigorously; for negative valence and positive arousal they might be fighting violently; for positive valence and negative arousal they might simply be sitting calmly; whilst for negative valence and arousal one might be on her deathbed, with the other in attendance. Of course, we were working with a relatively small sample of artworks tagged with emotion descriptors. Any ideal situation would be hard to create given the constraint of a limited database.

The tagged database of musical extracts at our disposal was even more limited, and again made it difficult to avoid confounding factors. For example, in an ideal situation we should have avoided music with lyrics. There was an additional problem for the music chosen by expert, in that we only had one person's assessment, rather than the minimum ten assessments for music extracts annotated by crowd-sourcing. We could have compensated for this by asking participants to assess the valence and arousal of each these musical extracts, either directly or by associating emotion tags. We could go further by asking participants to emotionally assess all the musical extracts, and the artworks. This would enable a somewhat different analysis. Additionally to analysing how people react to artwork and music positioned in the VA plane by a annotators, we could analyse how people react to artwork and music, in the light of their own positioning of the artwork and music. Participants' responses will be determined by their own view of the artwork and music, not by an average view of a group of other people.

We need also to consider the question of what we mean by valence, in particular in the context of artwork. Figure 1, taken from Russell (1980), uses the terms *misery* and *pleasure* to describe the horizontal axis; elsewhere in the paper he uses *pleasantness* to describe this axis. We have already noted that *valence* came to be used later. However, there are two interpretations of pleasantness, based on content or based on presentation. If we consider *The Hippopotamus and the Crocodile Hunt*, which we took to represent negative valence, the negativity seems likely to arise from the content matter, not the presentation; for the picture might be regarded as a glorious riot of colour. Another example is the picture we could have used to represent negative valence and positive arousal, but rejected because of its sensitive subject: *The Rape of Tamar*, by Le Sueur, which makes use of "harmonious" colours[28]. Our calculation of valence was based on the annotations provided by Mohammad and Kiritchenko (2018). Their instructions to annotators were: *"Examine the art above (the image). Which of the following describe the emotions it brings to mind? Select all that apply."* They also presented four examples of how emotions might be represented in any artwork. Three of these examples related to representative art, and were clearly aimed at content. Only the fourth, relating to non-representative art, was necessarily unrelated to content. Their annotators were provided with the tags which we show in Table 20. The majority of these tags relate to content, rather than presentation; with the possible exceptions of *agreeableness* and *disagreeableness*. Consequently our allocation of valence is likely to primarily reflect artwork content. When we ourselves asked participants to rate the pleasantness of an artwork, we gave no advice as to whether to concentrate on content or presentation. It would be useful in a future study to ask for two pleasantness ratings, one based on content and the other on presentation.

---
[28] See https://www.metmuseum.org/art/collection/search/436858

Table 20. List of emotions provided to annotators to label images; adapted from Mohammad and Kiritchenko (2018), Table 3.

| Polarity | Emotion category |
|---|---|
| Positive | **gratitude**, thankfulness, or indebtedness<br>**happiness**, calmness, pleasure, or ecstasy<br>**humility**, modesty, unpretentiousness, or simplicity<br>**love** or affection<br>**optimism**, hopefulness, or confidence<br>**trust**, admiration, respect, dignity, or honor |
| Negative | **anger**, annoyance, or rage<br>**arrogance**, vanity, hubris, or conceit<br>**disgust**, dislike, indifference, or hate<br>**fear**, anxiety, vulnerability, or terror<br>**pessimism**, cynicism, or lack of confidence<br>**regret**, guilt, or remorse<br>**sadness**, pensiveness, loneliness, or grief<br>**shame**, humiliation, or disgrace |
| Other or mixed | **agreeableness**, acceptance, submission, or compliance<br>**anticipation**, interest, curiosity, suspicion, or vigilance<br>**disagreeableness**, defiance, conflict, or strife<br>**surprise**, surrealism, amazement, or confusion<br>**shyness**, self-consciousness, reserve, or reticence<br>**neutral** |

For music, unless there are lyrics, there is no content, and emotional judgement must be based on the artform itself. Of course, participants might be influenced by the music's title; the title *Threnody to the Victims of Hiroshima* is likely to have a negative effect on participants, and ideally they should not be aware of titles. The categories used by Aljanaki et al. (2016), based on those of Zentner et al. (2008), were: *amazement, solemnity, tenderness, nostalgia, calmness, power, joyful activation, tension, sadness*. There are overlaps between these categories and the categories of Table 20, e.g. they both contain *sadness*, and *amazement* in the Aljanaki et al. (2016) occurs under the category *surprise* in Table 20. However, the emotions provided by Mohammad and Kiritchenko (2018), and shown in Table 20, appear to represent a wider range of emotions than those of Aljanaki et al. (2016). The latter might be regarded as emotional states, indeed the authors use the term *affect state*; the former include emotions with more cognitive and social implications, e.g. *arrogance* and *shame*. There is clearly much more that could be said on this topic, e.g. the OCC model of emotion (Ortony et al., 2022), which takes account of the role of cognition in creating emotions, may be relevant to this discussion. Our purpose here is merely to note that future studies in cross-modality, which take account of the role of emotion, need to consider the potential differences in kinds of emotions between modalities, and the extent to which the study is concerned with content as well as the essential artform.

In calculating the sentiment of participant's responses we needed to go beyond our approach for the artwork and music tags, and take account of the fact that the emotion words in the responses were embedded within sentences. Our approach for calculating valence and arousal was relatively simplistic. In particular, we compensated only for the effect of negators.

Kiritchenko and Mohammad (2017) have investigated the effect of negators, modals and degree adverbs, and their work would better enable compensating for the effect of words surrounding the lexicon words. Rathje et al. (2024) have investigated the use of GPT for text analysis, e.g. sentiment analysis, and claim to have achieved results superior to the use of lexicons. Their comparison included the Emotion Lexicon from the National Research Council of Canada (Mohammad & Turney, 2013), which categorizes words into eight emotions, plus positive or negative sentiment. Rathje et al. (2024) did not investigate the use of GPT to specifically determine the valence or arousal of text and this would be a valuable approach in future studies.

### 4.2 Key findings

In the following three subsections, we address our first three research questions. In subsection 4.2.4 we discuss some aspects of our findings which were not directly anticipated by our research questions.

#### 4.2.1 Cross-modal experience and the appreciation of art

Our first research question was about the effect of the music on art appreciation:
RQ1 Can cross-modal experiences enhance the appreciation of visual art?

It is clear that music does alter the experience of looking at art, although in varied ways. As noted in subsection 3.1.3, overall both groups of music depressed the perceived pleasantness of the artworks. Considering the music-artwork combinations individually, then for the music chosen by algorithm there were five combinations where the effect was significant; for the music chosen by expert, there were eight combinations where the effect was significant. For the music chosen by algorithm, all the significant effects were to depress the pleasantness of the artwork. For the music chosen by expert, six of the significant effects depressed the pleasantness, whilst two enhanced it.

When we consider the effect on meaningfulness, we find that overall only the music chosen by algorithm had a significant effect, and that was to depress the meaningfulness. Considering the music-artwork combinations individually, for the music chosen by algorithm there were ten combinations where the effect was significant, and for the music chosen by expert there were three combinations where the effect was significant. All of these significant effects depressed the meaningfulness of the artwork.

However, as the comments in subsection 3.2 show, these statistics mask a variety of experiences. It is clear that, for some participants, the music altered the experience of the art in a positive way, e.g. by providing a sense of motion, or by helping to maintain engagement with a painting for longer.

#### 4.2.2 Connections between music and art

Our second research question was about the connections participants made between the art and the music:
RQ2 What connections do people make between the art and the music during the appreciation of the visual art?

When we asked what connections participants were making between artwork and music, emotion was the most popular response. However, theme and story were also frequently cited. Movement was less frequently cited, but still occurred in around 15% of the responses. When asked to explain their choice of connections, a number of participants described themes or stories which they felt connected artwork and music; the stories, in particular, were sometimes quite imaginative. Participants also described how the music helped induce a sense of movement in the artwork. Sometimes this was relatively predictable, e.g. the quote in subsection 3.2.4.3 relating to Festivities on the Coast; at other times less so, e.g. the quote in the same section relating to The Hippopotamus and Crocodile Hunt. The latter quote occurred with music with positive valence and positive arousal, and the remark at the end of the quote, "turning it into something happy", suggests the effect of the emotion in the music. It seems here that the positive emotion of the music overcame the negativity of the artwork.

It was hard to discern from participants' responses what other connections they were making. Often, participants seem to have responded with *other* when one of the four options provided would have been appropriate. Participants responses when they felt there was no connection at all give an interesting insight. Some participants simply found the artwork and music incongruous, e.g. because of the lyrics, or because of a sense of anachronism. Perhaps *congruity* might be proposed as an additional form of connection; or perhaps congruity is a requirement to enable other forms of connection to work effectively. By *congruity* we mean something different from the congruence between emotions which we discuss in the next subsection.

### 4.2.3 Effects of interacting emotions

Our third research question was concerned with the interaction between the emotions conveyed by the music and the art:

    RQ3    How is the relationship between art and music affected by the emotions they individually conveyed, and by the congruence and incongruence of those emotions?

It was clear from subsection 3.1 that no simple relationship exists between the position of the artwork and music in the VA plane and the meaningfulness or pleasantness of the artwork, or the connectivity between artwork and music. Moreover, the different patterns which we see when we compare the effect of the music chosen by algorithm and chosen by expert suggest that there are other factors besides valence and arousal affecting the experience. An initial hypothesis that there would be a greater sense of connection between music and artwork when they were in the same quadrant of the VA plane was rejected. There was no significant difference between the four possible relationships between the music and artwork quadrants, as listed in subsection 3.1.1, i.e. same quadrant, opposing quadrants etc.

### 4.2.4 Additional findings

One unexpected finding relates to the effect of repetitive music, specifically in our study *Dance Pieces No. 9* and *Spiegel im Spiegel*. With this music, some participants reported an apparent slowing down of time. This seemed to be accompanied by an ability to *"calmly observe ... like someone who is there but who isn't there."*

A second comment relates to method. The application of sentiment analysis to the responses to the open-ended question did give further insight. This was particularly the case with the question about helpfulness, where it was possible to discern a statistically significant difference between the reactions to the artworks with positive and with negative arousal.

## 4.3 Guidelines for using cross-modality

Our fourth research question was practical:

RQ4  What guidance can be given on the use of music to enhance the experience of visual art?

It is clear that different people respond in different ways to the experience of viewing art and concurrently listening to music. Reactions ranged from participants who were very enthusiastic about the experience to at least one participant who expressed a preference for viewing in silence. Nevertheless, we are able to offer a few guidelines:
- An emotional match between the artwork and the music may be helpful. However, this is not always the case. For example, positively valenced music can be used to lift the mood of a negatively valenced artwork; and incongruity between music and artwork emotions might be used to create an unusual, quirky effect.
- Some forms of incongruity should be avoided, e.g. anachronisms; again, an exception might be if you were looking for a quirky effect.
- The use of repetitive music may, for some viewers, cause an apparent slowing down of time; this may encourage a calmer observation of the artwork.
- A shared theme or story can help make a connection between artwork and music. The experience of viewing and listening can be enhanced by encouraging the development of such themes or stories.
- It is likely that a better effect will be obtained when an expert chooses the music, rather than it being chosen by algorithm. In particular, an expert can take account of the range of factors which influence the experience, e.g. emotion, rhythm, shared theme or story.
- Given the wide range of reactions to the same experience, it may be useful to give viewers a choice of music provided individually on earphones.

## 4.4 Future research

Our study gives rise to some directions for future research:
- Further investigation of how varying emotions in art and music affect the combined experience. This needs to build on the study reported here, but taking account of our comments in subsection 4.1 on experimental design.
- Investigation of how people create stories when viewing art, how those stories enhance the experience, and how music can be used to stimulate story creation.
- Investigation of additional factors in music which might influence interaction with art, e.g. tempo and genre.
- Use of more refined techniques for emotion detection in text; this could extend to the more complex emotions described by Ortony et al. (2022). It would also be interesting to determine whether responses to art can be analysed using the aesthetic taxonomies developed by Christensen et al. (2023); or to gauge the stage of an individual's aesthetic understanding, e.g. on the scale developed by Housen (1983).

# 5 Conclusions

Our study has investigated the effect of music on the experience of viewing art, specifically the meaningfulness and pleasantness of the art, and the sense of connectivity between art and music. We used one set of four artworks and two sets of four musical extracts. One set of music was chosen by algorithm based on crowdsourcing data; the other was chosen by an expert. Each participant viewed one artwork and listened to one set of musical extracts. We found that music was capable of having a considerable effect on the experience. However, there was no simple relationship between the emotion conveyed by the music and the effect on the participants' experience. When asked how they formed connections between the artwork and the music, participants cited emotional connections, the awareness of shared themes, the development of stories; and, to a lesser extent, the tempo of the music creating a sense of movement in the art. Overall, participants found the music helpful in developing an appreciation of the artwork, with the music chosen by expert significantly more helpful than that chosen based on the crowdsourcing data.

Participants were also asked a number of open-ended questions to provide further elaboration. Answers to these open-ended questions provided further insight into how participants made connections between artwork and music. One unanticipated effect which participants reported, was that of an apparent slowing down of time when listening to certain repetitive pieces of music. This seemed to encourage a calm observation of the art. Text-mining was also used to assess the sentiment of the answers to the open-ended questions. This revealed complex relationships between the sentiment in the art and music, and the sentiment in the answers to these questions. However, there was a significant positive correlation between valence of participants' answers to questions about the artworks, prior to listening to music, and the reported pleasantness of the art. Also, the valence of the answers to questions about the helpfulness of the music was significantly higher for the two artworks with higher arousal than for the other two artworks, suggesting the music was more helpful in appreciating these artworks.

We have also provided some guidelines for using music to enhance the experience of looking at art. Given the wide range of reactions to our study, there is a strong argument for giving viewers a choice of music.

Finally, we provide some directions for future research. These include further investigation in the research questions discussed here, e.g. relating to the role of emotion, theme, story and tempo in cross-modality. They also include the use of more sophisticated text-mining techniques, not only for emotion detection, but for additional classification of people's responses to art.

## Acknowledgements

The authors would like to thank all those who gave up their time to participate in this study.

## Funding

The research has received funding from the European Union's Horizon 2020 research and innovation programme through the project Polifonia: a digital harmoniser of musical cultural heritage (Grant Agreement N. 101004746, https://polifonia-project.eu).

## References

Albertazzi, L., Canal, L., & Micciolo, R. (2015). Cross-modal associations between materic painting and classical Spanish music. *Frontiers in Psychology*, *6*, 424.

Aljanaki, A., Wiering, F., & Veltkamp, R. C. (2016). Studying emotion induced by music through a crowdsourcing game. *Information Processing & Management*, *52*(1), 115–128.

Christensen, A. P., Cardillo, E. R., & Chatterjee, A. (2023). What kind of impacts can artwork have on viewers? Establishing a taxonomy for aesthetic impacts. *British Journal of Psychology*, *114*(2), 335–351. https://doi.org/10.1111/bjop.12623

Cowles, J. T. (1935). An experimental study of the pairing of certain auditory and visual stimuli. *Journal of Experimental Psychology*, *18*(4), 461.

DeSantis, K., & Housen, A. (2009). *A brief guide to developmental theory and aesthetic development*. Visual Understanding in Education New York, NY.

Ferraris, C., Davis, T., Gatzidis, C., & Hargood, C. (2023). Digital Cultural Items in Space: The Impact of Contextual Information on Presenting Digital Cultural Items. *ACM Journal on Computing and Cultural Heritage*.

Fosh, L., Benford, S., Reeves, S., Koleva, B., & Brundell, P. (2013). see me, feel me, touch me, hear me: Trajectories and interpretation in a sculpture garden. *Proceedings of the SIGCHI Conference on Human Factors in Computing Systems*, 149–158. https://doi.org/10.1145/2470654.2470675

Hazzard, A., Benford, S., & Burnett, G. (2015). Sculpting a Mobile Musical Soundtrack. *Proceedings of the 33rd Annual ACM Conference on Human Factors in Computing Systems*, 387–396. https://doi.org/10.1145/2702123.2702236

Housen, A. (1983). *The eye of the beholder: Measuring aesthetic development*. Harvard University.

Housen, A. (1987). Three methods for understanding museum audiences. *Museum Studies Journal*, *2*(4), 41–49.

Iosifyan, M., Sidoroff-Dorso, A., & Wolfe, J. (2022). Cross-modal associations between paintings and sounds: Effects of embodiment. *Perception*, *51*(12), 871–888. https://doi.org/10.1177/03010066221126452

Isaacson, A., Assis, A., & Adi-Japha, E. (2023). "Listening" to Paintings: Synergetic Effect of a Cross-Modal Experience on Subjective Perception. *Empirical Studies of the Arts*, *41*(2), 433–464. https://doi.org/10.1177/02762374231155742

Janzen, T. B., de Oliveira, B., Ferreira, G. V., Sato, J. R., Feitosa-Santana, C., & Vanzella, P. (2023). The effect of background music on the aesthetic experience of a visual artwork in a naturalistic environment. *Psychology of Music*, *51*(1), 16–32.

Kiritchenko, S., & Mohammad, S. M. (2017). *The Effect of Negators, Modals, and Degree Adverbs on Sentiment Composition* (arXiv:1712.01794). arXiv. http://arxiv.org/abs/1712.01794

Leder, H., Belke, B., Oeberst, A., & Augustin, D. (2004). A model of aesthetic appreciation and aesthetic judgments. *British Journal of Psychology*, *95*(4), 489–508. https://doi.org/10.1348/0007126042369811

Leder, H., Carbon, C.-C., & Ripsas, A.-L. (2006). Entitling art: Influence of title information on understanding and appreciation of paintings. *Acta Psychologica*, *121*(2), 176–198.

Leder, H., & Nadal, M. (2014). Ten years of a model of aesthetic appreciation and aesthetic judgments: The aesthetic episode – Developments and challenges in empirical aesthetics. *British Journal of Psychology*, *105*(4), 443–464. https://doi.org/10.1111/bjop.12084

Limbert, W. M., & Polzella, D. J. (1998). Effects of music on the perception of paintings. *Empirical Studies of the Arts*, *16*(1), 33–39.

Mohammad, S. (2018). Obtaining reliable human ratings of valence, arousal, and dominance for 20,000 English words. *Proceedings of the 56th Annual Meeting of the Association for Computational Linguistics (Volume 1: Long Papers)*, 174–184.

Mohammad, S., & Kiritchenko, S. (2018). Wikiart emotions: An annotated dataset of emotions evoked by art. *Proceedings of the Eleventh International Conference on Language Resources and Evaluation (LREC 2018)*.

Mohammad, S. M., & Turney, P. D. (2013). NRC emotion lexicon. *National Research Council, Canada*, *2*, 234.

Mulholland, P., Stoneman, A., Barker, N., Maguire, M., Carvalho, J., Daga, E., & Warren, P. (2023). The Sound of Paintings: Using Citizen Curation to Explore the Cross-Modal Personalization of Museum Experiences. *Adjunct Proceedings of the 31st ACM Conference on User Modeling, Adaptation and Personalization*, 408–418.

Ortony, A., Clore, G. L., & Collins, A. (2022). *The cognitive structure of emotions*. Cambridge university press. https://books.google.co.uk/books?hl=en&lr=&id=1dd4EAAAQBAJ&oi=fnd&pg=PR9&dq=the+cognitive+structure+of+emotions&ots=aZAg3ySAg3&sig=3dr-gwM6_ZahxGOcOFRZe46rxOI

Özger, C., & Choudhury, N. (2023). Wikipedia and Shostakovich Meets Goya: Elaborative Narration and Music Enhance Affect Derived From Art. *Empirical Studies of the Arts*, 02762374231170260.

Palmer, S. E., Schloss, K. B., Xu, Z., & Prado-León, L. R. (2013). Music–color associations are mediated by emotion. *Proceedings of the National Academy of Sciences*, *110*(22), 8836–8841.

Park, J., Kim, M., & Kim, H. Y. (2024). Image Is All for Music Retrieval: Interactive Music Retrieval System Using Images with Mood and Theme Attributes. *International Journal of Human–Computer Interaction*, *40*(14), 3841–3855. https://doi.org/10.1080/10447318.2023.2201557

Rathje, S., Mirea, D.-M., Sucholutsky, I., Marjieh, R., Robertson, C. E., & Van Bavel, J. J. (2024). GPT is an effective tool for multilingual psychological text analysis. *Proceedings of the National Academy of Sciences*, *121*(34), e2308950121. https://doi.org/10.1073/pnas.2308950121

Russell, J. A. (1980). A circumplex model of affect. *Journal of Personality and Social Psychology*, *39*(6), 1161.

Russell, J. A., & Mehrabian, A. (1977). Evidence for a three-factor theory of emotions. *Journal of Research in Personality*, *11*(3), 273–294.

Russell, P. A. (2003). Effort after meaning and the hedonic value of paintings. *British Journal of Psychology*, *94*(1), 99–110.

Spence, C. (2020a). Assessing the role of emotional mediation in explaining crossmodal correspondences involving musical stimuli. *Multisensory Research*, *33*(1), 1–29.

Spence, C. (2020b). Simple and complex crossmodal correspondences involving audition. *Acoustical Science and Technology*, *41*(1), 6–12.

Swayne. (2024). *CRY (1979)*. https://www.wisemusicclassical.com/work/7888/CRY--Giles-Swayne/

Whiteford, K. L., Schloss, K. B., Helwig, N. E., & Palmer, S. E. (2018). Color, Music, and Emotion: Bach to the Blues. *I-Perception*, *9*(6), 204166951880853. https://doi.org/10.1177/2041669518808535

Zentner, M., Grandjean, D., & Scherer, K. R. (2008). Emotions evoked by the sound of music: Characterization, classification, and measurement. *Emotion*, *8*(4), 494.

Zhang, Z. (2018). Text Mining for Social and Behavioral Research Using R. *A Case Study on Teaching Evaluation. https://Books. Psychstat. Org/Textmining/Index. Html*.